\documentclass[useAMS,usenatbib]{mn2e}

\usepackage{graphicx,natbib, bm, amssymb, subfigure, amsmath, longtable}

\title[Stokes tomography of radio pulsar magnetospheres. II. Millisecond pulsars]{Stokes tomography of radio pulsar magnetospheres. II. Millisecond pulsars}

\author[C. T. Y. Chung et al.]{C. T. Y.~Chung $^1$\thanks{E-mail: cchung@physics.unimelb.edu.au} and A.~Melatos $^1$ \\
 $^1$ School of Physics, University of Melbourne, Parkville, VIC               
 3010, Australia}

\begin{document}

\date{ }

\pagerange{\pageref{firstpage}--\pageref{lastpage}} \pubyear{2011}

\maketitle

\label{firstpage}

\begin{abstract}
The radio polarization characteristics of millisecond pulsars (MSPs) differ significantly from those of non-recycled pulsars. In particular, the position angle (PA) swings of many MSPs deviate from the S-shape predicted by the rotating vector model, even after relativistic aberration is accounted for, indicating that they have non-dipolar magnetic geometries, likely due to a history of accretion. Stokes tomography uses phase portraits of the Stokes parameters as a diagnostic tool to infer a pulsar's magnetic geometry and orientation. This paper applies Stokes tomography to MSPs, generalizing the technique to handle interpulse emission. We present an atlas of look-up tables for the Stokes phase portraits and PA swings of MSPs with current-modified dipole fields, filled core and hollow cone beams, and two empirical linear polarization models. We compare our look-up tables to data from 15 MSPs and find that the Stokes phase portraits for a current-modified dipole approximately match several MSPs whose PA swings are flat or irregular and cannot be reconciled with the standard axisymmetric rotating vector model. PSR J1939+2134 and PSR J0437$-$4715 are modelled in detail. The data from PSR J1939+2134 at 0.61\,GHz  can be fitted well with a current-modified dipole at $(\alpha, i) = (22 \pm 2^\circ, 80 \pm 1^\circ)$ and emission altitude 0.4 $r_\text{LC}$. The fit is less accurate for PSR J1939+2134 at 1.414\,GHz, and for PSR J0437$-$4715 at 1.44\,GHz, indicating that these objects may have a more complicated magnetic field geometry, such as a localized surface anomaly or a polar magnetic mountain. 
\end{abstract}

\begin{keywords}
magnetic fields --- polarization --- pulsars: general --- pulsars: individual: PSR J0437$-$4715 --- PSR J1939+2134
\end{keywords}

\section{Introduction}
\label{sec:intro}

The two tools used most frequently to characterise the orientation and magnetic geometry of a radio pulsar are its pulse profile and position angle (PA) swing. The rotating vector model \citep{radha69}, which assumes an axisymmetric magnetic field, predicts an S-shaped swing across one pulse period and is traditionally used to determine the inclinations of the magnetic axis of symmetry and the observer's line of sight to the rotation axis. However, limitations arise when analysing only the PA swing, especially as the magnetosphere is not axisymmetric in general, e.g. the magnetic field includes a current-modified component \citep{hibschman01}. 

In \citet{chung10} (hereafter CM10), Stokes tomography was introduced as a diagnostic tool to be used alongside more traditional methods of analysis. It exploits the fact that the phase portraits traced out by the four Stokes parameters, when plotted against each other over one pulse period, are unique for any given magnetic geometry and orientation. An atlas of look-up tables, containing Stokes phase portraits and PA swings, was generated by CM10 for a variety of simple models, including pure and current-modified dipole fields, filled core and hollow cone beams, and the associated linear polarization patterns. CM10 also showed that, from a sample of 24 nominally ``dipolar" pulsars, which obey the period-pulse-width relation and/or exhibit clean S-shaped PA swings, the Stokes phase portraits of 16 objects are either inconsistent with low-altitude emission from a pure dipole field, or have a highly asymmetric surface emission pattern.

In this paper, we turn our attention to millisecond pulsars (MSPs). Polarimetric studies of MSP radio emission have uncovered complex behaviour not normally seen in slower pulsars. In particular, the PA swings of many MSPs are neither clean nor S-shaped; instead, they are flat \citep{stairs99, ord04}, highly distorted \citep[e.g. PSR J0437$-$4715;][]{navarro97} or extremely noisy \citep[e.g. giant pulses from PSR J1824$-$2452A;][]{knight06}. Additionally, the PA swing varies dramatically with frequency for many MSPs \citep{johnston08}, indicating that the magnetic geometry changes a lot with altitude, or that the observed pulse profile comprises emission from several different regions and altitudes.

The above trends suggest that MSPs have nondipolar magnetic fields. In a non-recycled pulsar, a dipole field can be distorted by several mechanisms, e.g. a current flowing along the field lines \citep{hibschman01, dyks08}, or rotational sweepback near the light cylinder \citep{hibschman01, dyks04, dyks08}. In a recycled pulsar with a history of prolonged accretion, another set of mechanisms comes into play. For example, accreted material channeled onto the magnetic poles distorts the frozen-in magnetic field as it spreads towards the equator \citep{melatos01, pamel04, zhang06, vigmel08}. Quadrupolar magnetic fields, proposed to explain the X-ray light curves of Her X-1 \citep{shakura91}, can even be comparable to the dipolar component \citep{long08}. Multipole fields can also be generated near the inner edge of the partially diamagnetic accretion disk of an X-ray pulsar \citep{lai99}. Alternatively, as the pulsar is spun up by accretion, the magnetic pole drifts towards the rotation axis, dragged inward by the motion of superfluid vortices in the pulsar's core \citep{srinivasan90, ruderman91, cheng97, lamb09}.

In this paper, we apply Stokes tomography to millisecond pulsar data drawn from the European Pulsar Network's (EPN) online database. In Section \ref{sec:stokes}, we briefly review the fitting recipe for determining the optimal orientation and beam polarization patterns from observed pulse profiles and Stokes phase portraits. We also extend the model in CM10 to treat interpulse emission. We compare our improved look-up tables of Stokes phase portraits and PA swings to observations of 15 MSPs in Section \ref{sec:population} to identify general trends. We then conduct detailed modelling of PSR J1939+2134, which has a strong interpulse, and PSR J0437$-$4715, which has a pulse with multiple peaks, in Sections \ref{sec:j1939} and \ref{sec:j0437} respectively. Our findings are summarised in Section \ref{sec:conclusion}.

\section{Stokes tomography}
\label{sec:stokes}
\subsection{Radiation field}
\label{sec:radiation}
For the convenience of the reader, we begin by summarising briefly how to determine the emission point and hence the polarization state of the radiation as a function of pulse longitude, following the recipe laid out in Section 2 of CM10. Our notation and definitions copy CM10.

We define two reference frames, as in Figure 1 of CM10: the inertial frame, in which the observer is at rest, with axes ($\mathbf{e}_x, \mathbf{e}_y, \mathbf{e}_z$), and the body frame of the pulsar. The relative motion between the frames is computed by solving Euler's equations of motion (including precession in general but not in this paper). The line-of-sight vector $\mathbf{w}$ is chosen to lie in the $\mathbf{e}_y$-$\mathbf{e}_z$ plane, making an angle $i$ with $\mathbf{e}_z$. The rotation and magnetic axes lie along $\mathbf{e}_z$ and one of the body frame axes ($\mathbf{e}_3$) respectively, separated by an angle $\alpha$. We define a spherical polar grid $(r, \theta, \phi)$ in the body frame covering the region $x_\text{min} \leq r/r_\text{LC} \leq x_\text{max}$, $0 \leq \theta \leq \pi$, $0 \leq \phi \leq 2\pi$, with $64 \times 256 \times 128$ grid cells, where the line $\theta = 0$ lies along $\mathbf{e}_3$,  and $r_\text{LC} = c/\Omega$ is the light cylinder radius. In this paper, we take $x_\text{min} = 0.01$ and $x_\text{max} = 0.83$ to accomodate the relatively small magnetospheres (and hence emission altitudes) of MSPs.

Radiation from highly relativistic particles flowing along magnetospheric field lines is narrowly beamed. Hence, without relativistic aberration, the observed emission point $\mathbf{x}_0(t)$ at any time $t$ is located where the magnetic vector $\mathbf{B [x}_0(t), t]$ points along $\mathbf{w}$. When aberration is included, the emission point $\mathbf{x}_0 (t)$ at time $t$ satisfies the equation \citep{blaskiewicz91, dyks08}
\begin{equation}
\label{eqn:tangent}
\mathbf{w} = \frac{\pm \mathbf{t}  + \mathbf{\Omega} \times \mathbf{x}_0/c}{\lvert \pm \mathbf{t}  + \mathbf{\Omega} \times \mathbf{x}_0/c \rvert},
\end{equation}
where  $\mathbf{t} = \mathbf{B}[\mathbf{x}_0(t), t]/|\mathbf{B[x}_0(t), t]|$ is the unit tangent vector to the magnetic field at $\mathbf{x}_0(t)$, and $\mathbf{\Omega}$ is the angular velocity vector.
CM10 considered emission from only one pole for simplicity [i.e. $+\mathbf{t}$ in equation (\ref{eqn:tangent})], and hence ignored interpulse emission. In this paper, we include emission from both the north and south poles, requiring both $\pm \mathbf{t}$ terms to be retained in (\ref{eqn:tangent}). At every instant, we thus have four emission points which satisfy (\ref{eqn:tangent}), two in the hemisphere opposite the observer (which he cannot see), and two facing the observer, which we label $P_1$ and $P_2$. 
We search the grid at a fixed altitude $r_0$ to find $P_1$ and $P_2$; the locations of $P_1$ and $P_2$ change with time in both the body frame and the inertial frame. 
 The $\mathbf{\Omega} \times \mathbf{x}_0$ term in (\ref{eqn:tangent}) encodes the aberration effect, as in \citet{hibschman01}. It is correct to order $\mathcal{O} (r/r_\text{LC})$ and should be replaced by the full relativistic expression when warranted by confidence in the model and data. 

The Stokes parameters ($I, Q, U, V$) associated with the complex electric field vector $\mathbf{E}$ at $\mathbf{x}_0(t)$, which describe the polarization state, are defined as
\begin{eqnarray}
I &=& \lvert E_x \rvert^2 + \lvert E_y \rvert^2\\
\label{Qeqn} Q &=&\lvert E_x \rvert^2 - \lvert E_y \rvert^2\\
\label{Ueqn} U &=& 2 \text{Re} (E_x E_y^*)\\
V &=& 2 \text{Im} (E_x E_y^*),
\end{eqnarray}
where $I$ is the polarised fraction of the total intensity, $L = (Q^2 + U^2)^{1/2}$ is the linearly polarised component, and $V$ is the circularly polarised component. The observed electric field vector \textbf{E}, assumed to be in the direction of the particle acceleration\footnote{The instantaneous acceleration is inclined slightly with respect to the normal (or binormal) of $\mathbf{B}$ at $\mathbf{x}_0 (t)$, because the emitting charges corotate. For more details, see the discussion around equation (2) in CM10 and equation (A3) in \citet{dyks08}.}, is the incoherent sum of the electric field vectors at $P_1$ and $P_2$, viz. $\mathbf{E} = \mathbf{E}_1 + \mathbf{E}_2$, where the relative phase between $\mathbf{E}_1$ and $\mathbf{E}_2$ fluctuates randomly. The observed Stokes parameters therefore reduce to $I = I_1 + I_2, Q = Q_1 + Q_2,$ and $U = U_1 + U_2$. In this paper, we assume that all the emission is linearly polarised for simplicity, i.e. $V = 0$. Circular polarization will be examined in a companion paper.

The $x$- and $y$- components are measured with respect to an orthonormal basis $(\mathbf{\hat{x}}, \mathbf{\hat{y}})$ which is fixed in the plane of the sky. In this paper, we choose $\hat{\mathbf{x}} = \mathbf{\Omega}_\text{p}/\lvert \mathbf{\Omega}_\text{p}\rvert$ and $\mathbf{\hat{y}} = \hat{\mathbf{x}} \times \mathbf{w} $, where $\mathbf{\Omega}_\text{p} = \mathbf{\Omega - (\Omega \cdot w) w}$ is the projection of $\mathbf{\Omega}$ onto the sky.
Then the polarization angle, $\psi$, between $\hat{\mathbf{x}}$ and the linearly polarised part of $\mathbf{E}$ is given by 
\begin{equation}
\label{paeqn}
\psi = \frac{1}{2} \tan^{-1} \frac{U}{Q}.
\end{equation}

The observational data obtained from the EPN are not necessarily expressed in the canonical basis $(\mathbf{\hat{x}}, \mathbf{\hat{y}})$. However, the $Q$-$U$ phase portrait has the same shape in any Cartesian basis; if the basis is rotated by an angle $\beta$ with respect to $\mathbf{\hat{x}}$ and $\mathbf{\hat{y}}$, the $Q$-$U$ phase portrait rotates by an angle $2 \beta$ without being distorted, unlike the $I$-$Q$ and $I$-$U$ phase portraits, which change shape. Hence, when analysing the data, the first step is to reproduce the shape of the $Q$-$U$ phase portrait as closely as possible without worrying about the orientation; infer $\beta$; rotate the $(I, Q, U, V)$ data into the canonical basis provisionally defined through $\beta$; and then adjust $\alpha, i$, and the beam and polarization patterns \textit{iteratively} to reproduce the $I$-$Q$ and $I$-$U$ portraits. The recipe for doing so is explained in Section 2.6 and Figure 3 of CM10.

\subsection{Look-up tables of Stokes phase portraits}
Figures \ref{m0w10lcostheta_tqvsi}--\ref{m25w10lsintheta_tpa} in the Appendix display look-up tables of Stokes phase portraits and PA swings, similar to those in CM10, updated to include interpulse emission. The figures are organised into four groups, corresponding to two beam models (filled core and hollow cone) and two polarization models ($L \propto \cos \theta$, $L \propto \sin \theta$; see CM10). All the look-up tables are for a current-modified dipole magnetic field (CM10) composed of a pure dipole plus a toroidal component with magnitude
\begin{equation}
B_\phi = - B_p \cos \alpha \sin \theta r/r_\text{LC},
\end{equation}
where $B_p = (B_r^2 + B_\theta^2)^{1/2}$ is the poloidal field strength.

\subsection{Interpulses}
An interpulse is a secondary pulse separated from the main pulse by approximately $180^\circ$ of rotational phase \citep{manchester77}. It is believed to arise when a pulsar is a nearly orthogonal rotator viewed nearly side-on, shining from both magnetic poles, i.e. with $\alpha \approx i \approx 90^\circ$, where `$\approx$' means `within roughly one beam width' in this context \citep{petrova08}.

\begin{figure}
\includegraphics[scale=0.5]{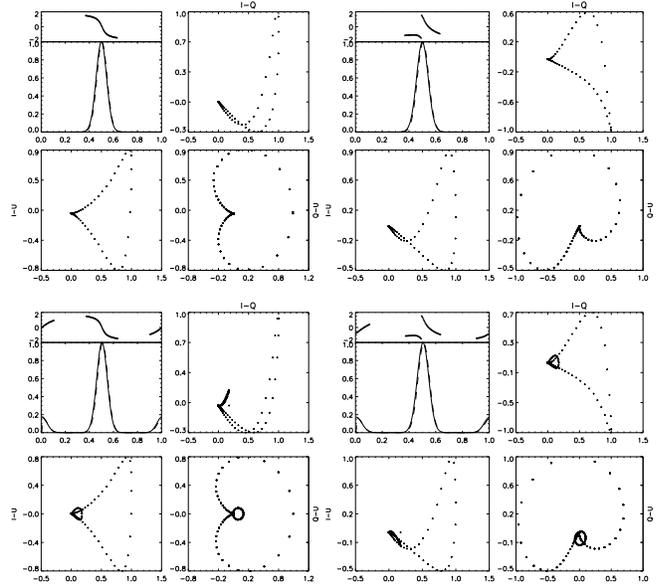}
\caption{Stokes tomography of a model pulsar with an interpulse but without relativistic aberration. Top left: dipole field, no interpulse. Top right: current-modified dipole emitting at $r = 0.13 r_\text{LC}$, no interpulse. Bottom left: dipole field with interpulse. Bottom right: current-modified dipole emitting at $r = 0.13 r_\text{LC}$ with interpulse. Within each quadrant of the figure, the five subpanels display (clockwise from top left): $I/I_\text{max}$ and PA (in radians) as functions of pulse longitude, $I$-$Q$, $Q$-$U$, and $I$-$U$. The orientation is $(\alpha, i) = (80^\circ, 70^\circ)$. The beam pattern is given by (\ref{eq:beam}).}
\label{compip}
\end{figure}

Figure \ref{compip} compares the Stokes phase portraits, pulse profiles, and PA swings for a pure dipole and current-modified dipole emitting from one and two poles for one illustrative orientation $(\alpha, i) = (80^\circ, 70^\circ)$. For clarity, relativistic aberration is \textit{not} included in this example (compare Section \ref{sec:aberration} \textit{et seq.}). Clockwise from the top left panel, the figure displays (i) a pure dipole with no interpulse, (ii) a current-modified dipole at $r = 0.13 r_\text{LC}$ with no interpulse, (iii) a pure dipole with an interpulse, and (iv) a current-modified dipole at $r = 0.13 r_\text{LC}$ with an interpulse. 
The top two panels in Figure \ref{compip}, which have no interpulse, are the same as in Figures 5--8 and 30--33 in the look-up tables in CM10.

\begin{figure}
\includegraphics[scale=0.45]{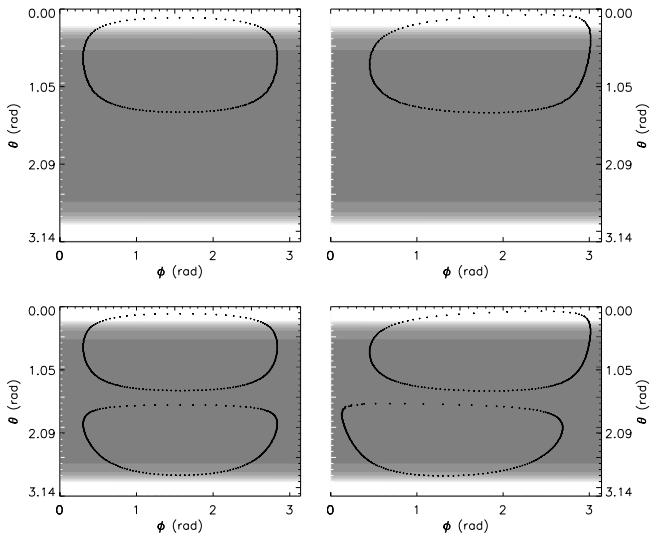}
\caption{Locus $\hat{\mathbf{x}}_0(t)$ traced out by the emission point(s) $P_1$ and $P_2$ over one rotation, in the body frame of the model pulsar considered in Figure \ref{compip}. Top left: dipole field, no interpulse. Top right: current-modified dipole emitting at $r = 0.13 r_\text{LC}$, no interpulse. Bottom left: dipole field with interpulse. Bottom right: current-modified dipole emitting at $r = 0.13 r_\text{LC}$ with interpulse. The beam pattern $I(\theta, \phi)$ is represented by greyscale shading (brightness $\propto I$).}
\label{compip2}
\end{figure}

Figure \ref{compip2} shows the loci $\hat{\mathbf{x}}_0(t)$ traced out by $P_1$ and $P_2$ over one rotation in the body frame of the pulsar for the various cases in Figure \ref{compip}. The panels are arranged as in Figure \ref{compip}. The bottom panels, in which the interpulse is present, show two paths, one in the north hemisphere, and one in the south. For the current-modified dipole (right panels), the loci are asymmetric, as discussed in CM10.
For definiteness, we consider a filled-core beam, viz.
\begin{eqnarray}
\nonumber I(\theta, \phi) &=& (2 \pi \sigma^2)^{-1/2} \left\{ \text{exp}\left[-\theta^2/(2 \sigma^2)\right]  \right. \\
\label{eq:beam} &&\left. + \text{exp}\left[-(\theta - \pi)^2/(2 \sigma^2)\right] \right\},
\end{eqnarray}
which is represented by greyscale shading in Figure \ref{compip2}. In (\ref{eq:beam}), $\sigma$ is the width of the beam, chosen arbitrarily to equal $10^\circ$. We also choose the linear polarization pattern to be 
\begin{equation}
 L(\theta, \phi) = I(\theta, \phi) \lvert \cos \theta \rvert
\end{equation}
in Figures \ref{compip} and \ref{compip2}. Other choices (e.g. $L \propto \sin \theta$) are equally valid and have been found empirically by CM10 to match the observational data in many objects.

In Figure \ref{compip}, the interpulse traces out a small, secondary loop within the primary pattern in the $I$-$Q$, $I$-$U$ and $Q$-$U$ phase portraits. It also changes slightly the range of $U$ and $Q$ covered by the main pulse. For example, for the dipole (left panels), the maximum value of $U$ decreases from 0.9 to 0.8 with the addition of the interpulse. 
The size of the secondary loop (i.e. the intensity of the interpulse) increases relative to the main pulse as $\alpha$ approaches $i$, as expected. Along $\alpha = 90^\circ$ and $i = 90^\circ$, when the interpulse and the main pulse peak at the same intensity, the primary and secondary patterns overlap in the $I$-$Q$, $I$-$U$ and $Q$-$U$ planes, and the phase portraits are indistinguishable from the non-interpulse case. The shapes do not overlap exactly for other orientations, where the main pulse is brighter than the interpulse.

In the $I$-$U$ and $Q$-$U$ planes, the balloons and heart shapes seen in CM10 are also seen when an interpulse is present. For example, for $\alpha = i = 90^\circ$, the Stokes parameters trace out two reflection-symmetric patterns with positive and negative $U$ to form complex, interlocking shapes (see Figures \ref{m0w10lcostheta_tqvsi}--\ref{m25w10lsintheta_tuvsq} from the atlas of look-up tables in the Appendix). As expected, the patterns are more intricate for a hollow cone than for a filled core. For example, the $Q$-$U$ portrait at $(\alpha, i) = (70^\circ, 80^\circ)$ for a filled core contains an asymmetric, tilted heart shape and a small oval, both connected at $Q = U = 0$ (Figure \ref{m0w10lcostheta_tuvsq}). The same orientation for a hollow cone shows a broader heart shape with two large, secondary ovals (Figure \ref{m25w10lcostheta_tuvsq}).

\subsection{Relativistic aberration}
\label{sec:aberration}
In the observer's reference frame, charged particles flowing outwards ultra-relativistically along poloidal magnetic field lines also have a small transverse velocity component because they corotate with the star as part of the highly conducting magnetosphere. This displaces $\mathbf{x}_0(t)$ by a distance of order $r/r_\text{LC}$ compared to its position when aberration is neglected. The electric field vector (parallel to the particle's acceleration vector) is also displaced, resulting in the well-known delay-radius relation \citep{blaskiewicz91, hibschman01, dyks08}. According to this relation, the centre of the pulse profile leads the steepest point of the PA swing by $4 r/r_\text{LC}$. 

We compute $\mathbf{x}_0(t)$ directly, including aberration, by solving (\ref{eqn:tangent}) numerically. As a cross check, we compare the numerical solution with the analytic approximation given by equation (F2) of \citet{hibschman01}, where the tangent field at the aberration-shifted emission point, $\mathbf{t}$, can be expressed as the sum of the tangent field at the original, non-aberrated emission point, $\mathbf{t}_0$, plus a perturbation $\mathbf{t}_1 = (\mathbf{\Omega} \times \mathbf{x}_0)/c - [(\mathbf{\Omega} \times \mathbf{x}_0)/c \cdot \mathbf{t}_0] \mathbf{t}_0$. As the aberration-induced deflection angle grows linearly with $r$, equation (F2) holds most accurately for small $r$. At $r = 0.13 r_\text{LC}$ and $0.31 r_\text{LC}$, the direct and approximate calculations of $\mathbf{t}$ agree to within $\sim$10\% and $\sim$20\% respectively. Note that, although we calculate $\mathbf{t}$ directly from (\ref{eqn:tangent}), the expression $\pm \mathbf{t} + \mathbf{\Omega} \times \mathbf{x}_0/c$ in (\ref{eqn:tangent}) itself breaks down near the light cylinder, where quadratic relativistic corrections come into play.

\begin{figure}
\includegraphics[scale=0.5]{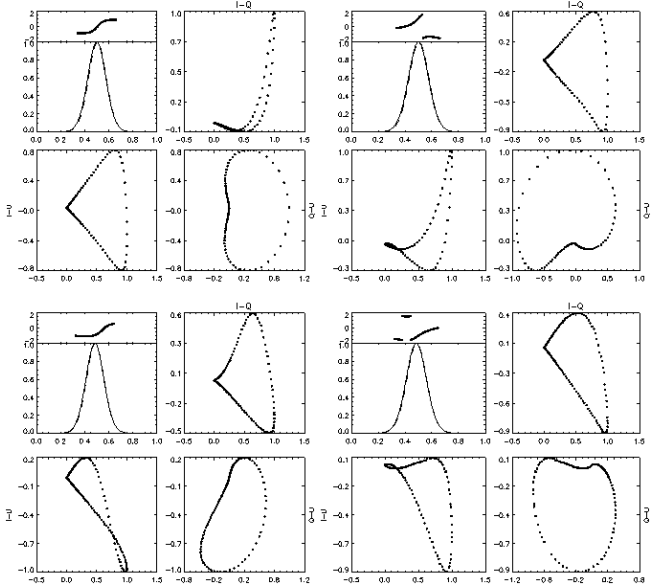}
\caption{Stokes tomography for a model pulsar including relativistic aberration but without an interpulse. Top left: dipole field, no aberration. Top right: current-modified dipole emitting at $r = 0.1 r_\text{LC}$, no aberration. Bottom left: dipole field emitting at $r = 0.1 r_\text{LC}$ with aberration. Bottom right: current-modified dipole emitting at $r = 0.1 r_\text{LC}$ with aberration. Within each quadrant of the figure, the five subpanels displays (clockwise from top left): $I/I_\text{max}$ and PA swing (in radians) as a function of pulse longitude, $I$-$Q$, $Q$-$U$, $I$-$U$. The orientation is $(\alpha, i) = (30^\circ, 40^\circ)$. The beam pattern is given by (\ref{eq:beam}).}
\label{compab}
\end{figure}

Figure \ref{compab} illustrates how aberration modifies the Stokes phase portraits, pulse profiles, and PA swings for pure dipole and current-modified dipole magnetospheres. For the sake of clarity, we do not include interpulse emission in Figure \ref{compab}, although, in general, interpulse and aberration effects are additive, as one can tell from the look-up tables in the Appendix. The emission is placed arbitrarily at an altitude of $0.1 r_\text{LC}$ to ensure a reasonably strong effect. We note that aberration introduces an altitude dependence in the case of a pure dipole, which is absent in the non-aberrated dipole considered in CM10. 

\begin{figure}
\includegraphics[scale=0.45]{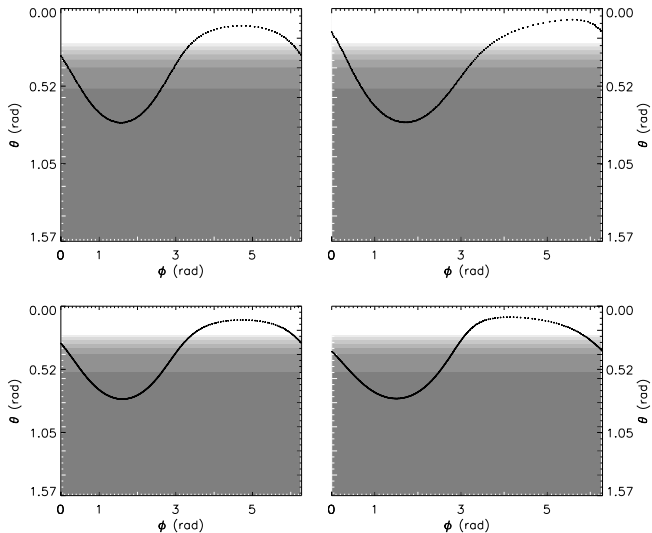}
\caption{Locus $\hat{\mathbf{x}}_0(t)$ traced out by the emission point $P_1$ across one rotation, in the body frame of the model pulsar considered in Figure \ref{compab}. Top left: dipole field, no aberration. Top right: current-modified dipole emitting at $r = 0.1 r_\text{LC}$, no aberration. Bottom left: dipole field emitting at $r = 0.1 r_\text{LC}$ with aberration. Bottom right: current-modified dipole emitting at $r = 0.1 r_\text{LC}$ with aberration. The beam pattern $I(\theta, \phi)$ is represented by greyscale shading (brightness $\propto I$).}
\label{compab2}
\end{figure}

Aberration acts mainly to shift the relative phases of the pulse centroid and PA swing inflection point. To lowest order in $r/r_\text{LC}$, the radius-delay relation predicts that the pulse profile is phase shifted by $\approx -r/r_\text{LC}$ radians, whereas the PA swing is phase shifted by $\approx 3 r/r_\text{LC}$ radians. Figure \ref{compab} shows that, for $r = 0.1r_\text{LC}$, the pulse profile is shifted by $\approx -$0.08 radians, whereas the PA swing is shifted by $\approx +$0.27 radians. These shifts are enough to dramatically broaden the $I$-$Q$ pattern, twist the $I$-$U$ pattern, and tilt the $Q$-$U$ pattern for the pure dipole (left panels of Figure \ref{compab}). For the current-modified dipole (right panels), the $I$-$Q$ pattern narrows, the $I$-$U$ patterns twists and rotates, and the $Q$-$U$ pattern rotates. Figure \ref{compab2} shows the loci of $\hat{\mathbf{x}}_0(t)$ traced out by $P_1$ over one rotation (there is only one set of emission points without an interpulse), with each panel corresponding to the cases in Figure \ref{compab}. The loci of the aberrated emission points (bottom panels) are shifted in $\phi$ relative to the non-aberrated points (top panels). 

In CM10, it is shown that, for a pure dipole field without aberration, all phase portraits are reflection symmetric about $U = 0$. Aberration breaks this symmetry, causing the shapes in the $I$-$U$ and $Q$-$U$ plane to tilt (see Section 4 in CM10). Aberration also changes the tilt and the relative sizes of the shapes in the phase portraits, e.g. the ventricles of the hearts in the $U$-$Q$ plane.

\subsection{Tilted axis of symmetry}
\begin{figure}
\centering
\includegraphics[scale=0.6]{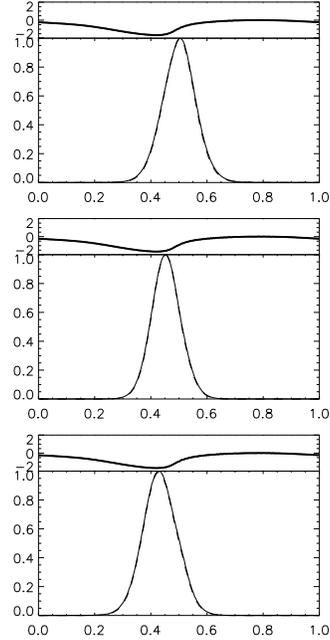}
\caption{Pulse profiles and PA swings for pure dipoles with identical orientations ($\alpha, i$) = ($30^\circ, 40^\circ)$. The left panels show beam axes offset by $(\theta', \phi')$ with respect to the magnetic axis, corresponding to ($\theta', \phi'$) = ($10^\circ, 0^\circ)$ (top), ($\theta',\phi'$) = $(10^\circ, 45^\circ$) (middle), and ($\theta',\phi'$) = ($10^\circ, 90^\circ$) (bottom). In the left panels, the emission radius is fixed at $r = 0.02 r_\text{LC}$. The right panels correspond to emission heights $r = 0.02 r_\text{LC}$ (top), $0.07 r_\text{LC}$ (middle), and $0.1 r_\text{LC}$ (bottom). In the right panels, the magnetic and beam axes are aligned.}
\label{plotoffset}
\end{figure}

If the beam pattern is centred on another axis that is slightly tilted with respect to the magnetic axis \citep[e.g. fan beams in the outer magnetosphere;][]{cheng00, watters09}, the pulse profile is also phase-shifted relative to the PA swing. Tilting the beam axis away from the magnetic axis can therefore mimic closely (though not exactly) the effects of aberration (see Section 2.4.1 in CM10). To illustrate, Figure \ref{plotoffset} compares the phase-shift caused by the tilt of the beam axis to that caused by aberration. We plot six pulse profiles and PA swings for a pure dipole with ($\alpha, i$) = ($30^\circ, 40^\circ$). The three panels on the left correspond to beam axes which are tilted with respect to the magnetic axis by $(\theta', \phi')$ = $(10^\circ, 0^\circ)$ (top), $(10^\circ, 45^\circ)$ (middle), and $(10^\circ, 90^\circ)$ (bottom), all emitting at $r = 0.02 r_\text{LC}$. Using the top panel as a reference point, the pulse centroid leads the PA swing by $-0.18$\,rad in the middle panel, and by $-0.36$\,rad in the bottom panel. On the right-hand side of Figure \ref{plotoffset}, the beam axis and magnetic axis are aligned, but we vary the emission radius. As aberration causes the pulse centroid to lead the PA swing by $-4 r/r_\text{LC}$, one could just as well attribute the phase shifts seen in the left-hand side to the emission radius increasing from $r = 0.02 r_\text{LC}$ (top), $0.07 r_\text{LC}$ (middle), and $0.1 r_\text{LC}$ (bottom). 

As it is not possible to distinguish the two effects without additional a priori information, throughout this paper we assume the beam is centred on the magnetic axis, except in Sections \ref{sec:j1939} and \ref{sec:j0437}, where we work with tilted beams reconstructed empirically from the pulse shape (because the magnetic-pole-centred model does not fit the data).

\section{Miniature population study}
\label{sec:population}

We now survey the pulse profiles, Stokes phase portraits, and PA swings for a selection of 16 MSPs from the EPN online database\footnote{Available at: http://www.mpifr-bonn.mpg.de/pulsar/data/} \citep{lorimer98}. These objects are chosen because they have pulse periods $< 10$\,ms, except for PSR J1022+1001, which was included because of its interesting Stokes phase portraits and because its orientation angles $\alpha$ and $i$ have been measured with some degree of confidence by previous authors \citep{stairs99}. All objects were observed by \citet{stairs99}, except for PSR J0437$-$4715 \citep{manchester95}. We note that the PA swing $\psi$ defined in this paper corresponds to $-\psi$ in \citet{stairs99}.
Table \ref{tab:pop} quotes the size of the magnetosphere for each MSP (in units of the stellar radius, $r_\star$) and the frequencies where EPN data are available. In the few cases where rotating vector model fits have been attempted in the literature, $\alpha$ and $i$ are also quoted, together with the published uncertainties.

\begin{table*}
 \centering
\label{tab:pop}
\caption{Pulse periods $P$ and observation frequencies of 15 millisecond pulsars from the EPN online database with $P < 10$\,ms (except for J1022+1001, which is included because of its interesting Stokes phase portraits). Where a rotating vector model fit has been published previously, $\alpha$ and $i$ are quoted with their published uncertainties. The uncertainties for $\alpha$ and $i$ for PSR J0437$-$4715 are not given in the literature. All pulsars were observed by \citet{stairs99}, except for PSR J0437$-$4715 \citep{manchester95}.}
\begin{tabular}{|l|r|r|c|l|l|}
\hline
Pulsar & $P$ (ms) & $r_\text{LC}/r_\star$ & Frequency (GHz) & $\alpha$ ($^\circ$) & $i$ ($^\circ$) \\
\hline
J0034$-$0534 & 1.88 & 9.09& 0.41 & &\\
J0218+4232 & 2.32 & 11.11 & 0.41 & $8  \pm 11$ & ---\\
		      & & & 0.61 & $8 \pm 15$ & ---\\
J0437$-$4715 & 5.76 & 27.78 & 1.44 & 145 & 140\\
		      & & & 4.6 & &\\
J0613$-$0200 & 3.06 & 14.71 & 0.41 & &\\
		      & & & 0.61 & &\\
J1012+5307 & 5.26 & 25.64 & 0.61 & &\\
J1022+1001 & 16.45 & 83.33 &  0.41 & &\\
		      & & & 0.61 & $140 \pm 16$ & $135.1 \pm 4.1$\\
		      & & & 1.414 & $83 \pm 27$ & $75.9 \pm 5.2$\\
J1643$-$1224 & 4.62 & 22.22 & 0.61 &&\\
J1713+0747 & 4.57 & 22.22 &  0.41 &&\\
		      & & & 0.61 &&\\
		      & & & 1.414 &&\\
J1730$-$2304 & 8.12 & 40.00 & 0.61 &&\\
J1744$-$1134 & 4.07 & 19.61 & 0.61 &&\\
J1823$-$3021A & 5.44 & 26.32 & 0.61&\\
J1824$-$2452 & 3.05 & 14.49 & 0.61& $40.7 \pm 1.7$ & $80.7 \pm 3.9$\\
J1911$-$1114 & 3.63 & 17.54 &  0.41 &&\\
		      & & & 0.61 &&\\
%J1932+1059 & 226.52 & 1086.96 &0.41 & $41 \pm 8$ & $62 \pm 9$\\
%		      & & & 0.61 & $51 \pm 3$ & $86 \pm 4$\\
%			& & & 1.414 & $61 \pm 2$ & $100 \pm 3$\\
J1939+2134 & 1.56 & 7.69 & 0.61 &&\\
		      & & & 1.414 &&\\
J2051$-$0827 & 4.51 & 21.74 & 0.41 &&\\
		      & & & 0.61 &&\\
\hline
\end{tabular}
\end{table*}

\subsection{General trends}
 In Figures \ref{data1}--\ref{data3}, we present the Stokes phase portraits, pulse profiles, and PA swings for the eight pulsars with the cleanest data. All Stokes parameters are normalized by the peak intensity $I_\text{max}$. MSPs generally have a lower degree of linear polarization than non-recycled pulsars, so their phase portraits are correspondingly noisier. The PA swing is only drawn at pulse longitudes satisfying $L \geq 0.1 L_\text{max}$ and $I \geq 0.1 I_\text{max}$, where $L_\text{max}$ is the peak value of $L$. As the absolute orientation of $\mathbf{\Omega}_p$ (and hence the angle $\beta$ between the measured and canonical bases) for each set of data is unknown, we start by analysing just the shape of the $Q$-$U$ phase portraits, as discussed in Section \ref{sec:radiation} and CM10.

Many of the pulse and linear polarization profiles are highly asymmetric, suggesting a complex emission pattern. In objects where $I(t)$ and $L(t)$ have multiple peaks or interpulse emission, the Stokes phase portraits feature multiple loops, each corresponding to an individual peak. For example, each of the five peaks seen in PSR J0437$-$4715 at 1.44\,GHz (Figure \ref{data1}, top row) corresponds to a distinct sub-pattern in the $I$-$Q$ and $I$-$U$ planes (see Section \ref{sec:j0437}). In the $Q$-$U$ plane, the pattern formed is an asymmetric figure-eight over a slightly curved line. PSR J1022+1001 (Figure \ref{data1}, rows 4--6) has an asymmetric, double-peaked pulse profile, which produces an asymmetric heart shape in the $Q$-$U$ plane. Also interesting is PSR J1939+2134 (Figure \ref{data3}, rows 4--5), which has a strong interpulse, whose phase portraits narrow with increasing frequency, while those of the main pulse broaden. 

The PA swings for the MSPs featured in Figures \ref{data1}--\ref{data3} are less informative. In several cases, where the PA swing is flat or noisy, the phase portraits still trace out a recognisable pattern. For example, in Figure \ref{data1}, the PA swing of PSR J0437$-$4715 is flat overall but punctuated by several dips, while the PA swing of PSR J1012+5307 is completely flat. Their phase portraits, in contrast, reveal balloons and figure-eights. In Figure \ref{data2}, the PA swings of PSR J1713+0747, PSR J1744$-$1134 and PSR J1911$-$1114 are flat, yet their phase portraits are distinguished by straight lines and balloons. The same is true of PSR J1939+2134 in Figure \ref{data3}. This is another instance, to be added to those in CM10, where the Stokes phase portraits carry important extra information which is not apparent from the pulse profile and PA swing alone.

\subsection{Magnetic geometry and orientation}
\begin{figure*}
\includegraphics[scale=0.8]{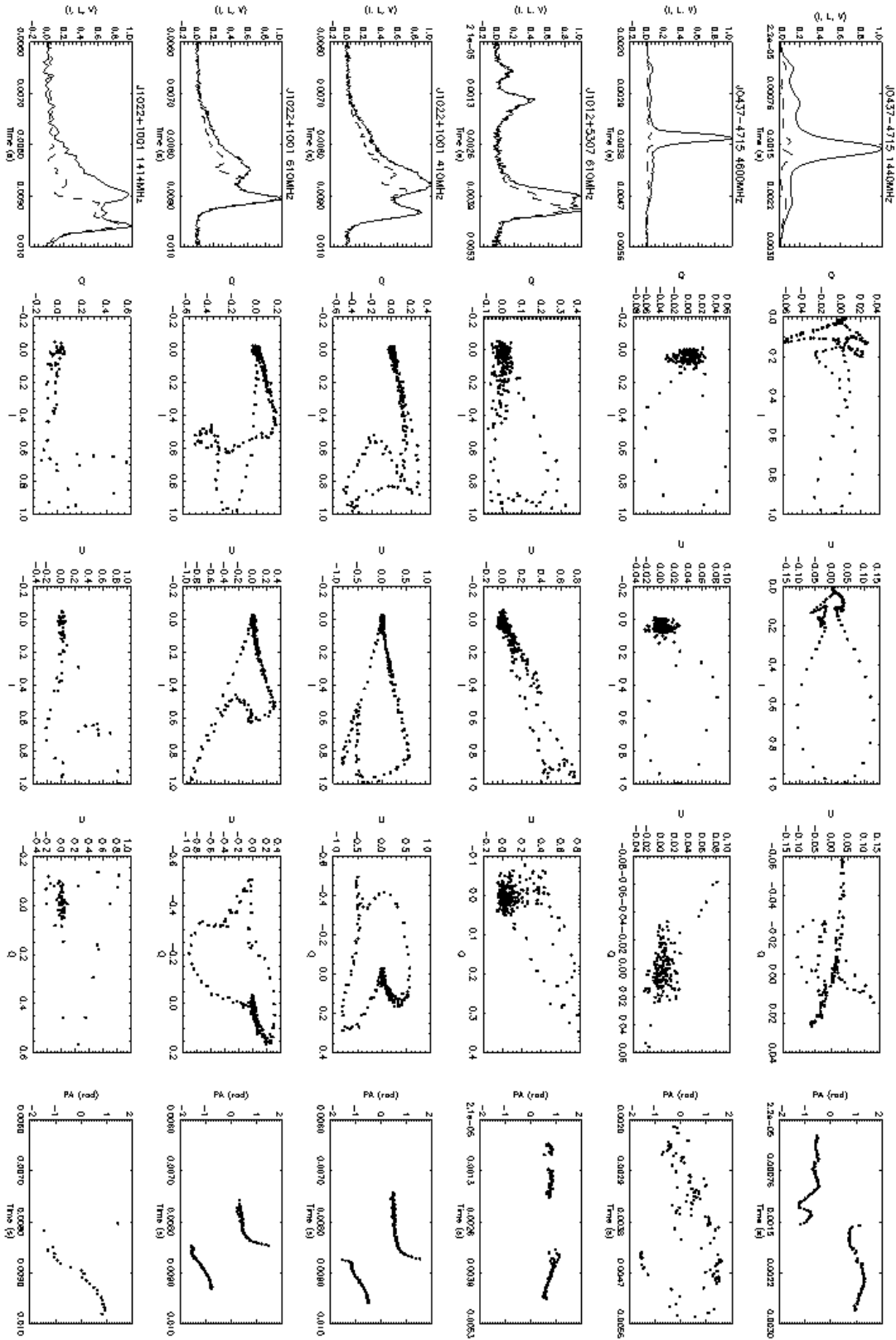}
\caption{Pulse profiles, Stokes phase portraits and PA swings for PSR J0437$-$4715 at 1.44\,GHz and 4.6\,GHz \citep{manchester95}, PSR J1012+5307 at 0.61\,GHz, and PSR J1022+1001 at 0.41\,GHz, 0.61\,GHz, and 1.414\,GHz \citep{stairs99}. The data for each pulsar occupy a row in landscape layout. From left to right, the columns show: (1) $I/I_\text{max}$ (solid curve) and $L/I_\text{max}$ (dashed curve) versus time (in s), (2) the $I$-$Q$ phase portrait, (3) the $I$-$U$ phase portrait, (4) the $Q$-$U$ phase portrait, (5) the PA swing  versus time (in s) (data points with $L \geq 0.1 L_\text{max}$ and $I \geq 0.1 I_\text{max}$ plotted only). Data are presented courtesy of the EPN online archive.}
\label{data1}
\end{figure*}

\begin{figure*}
\includegraphics[scale=0.8]{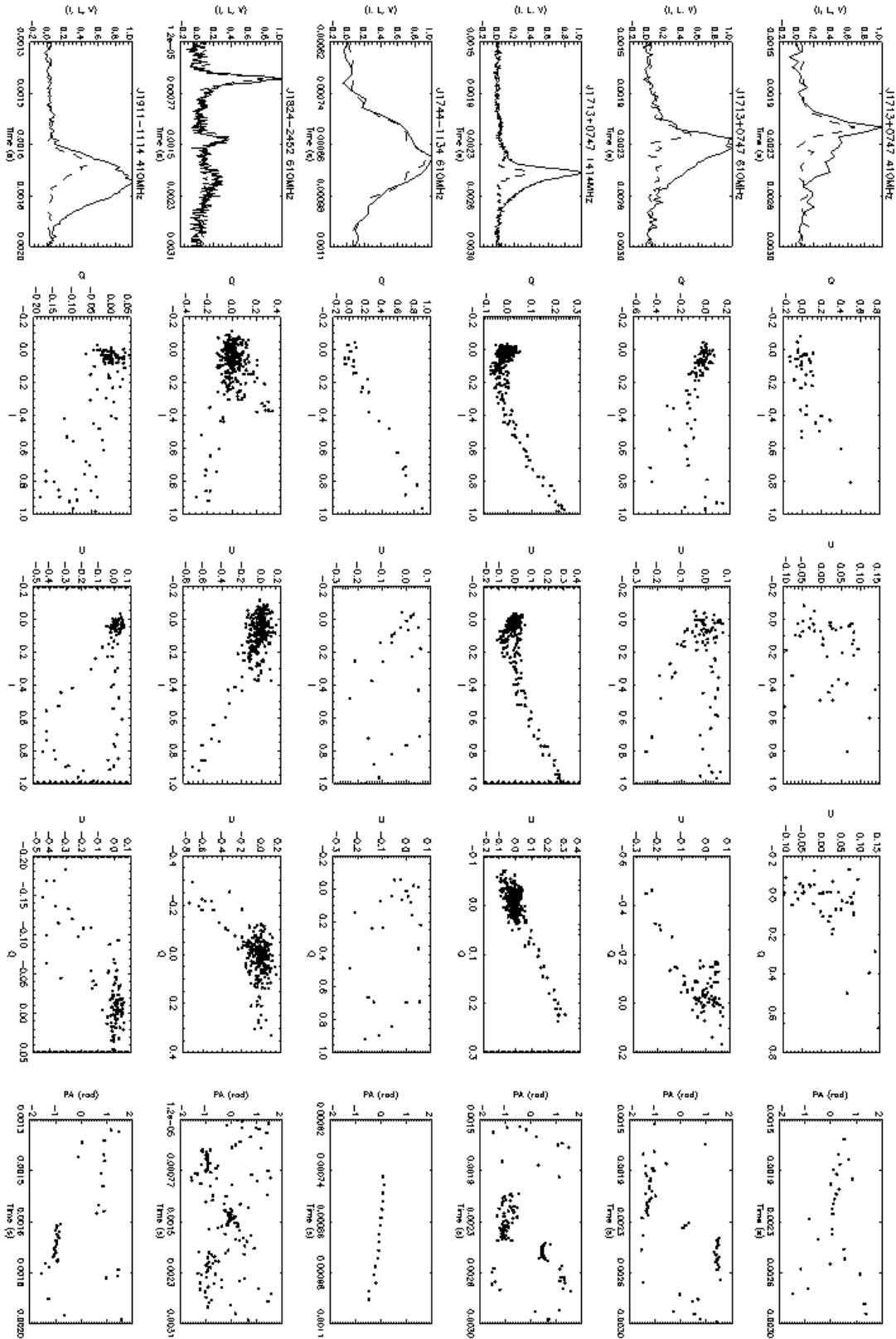}
\caption{Pulse profiles, Stokes phase portraits and PA swings for PSR J1713+0747 at 0.41\,GHz, 0.61\,GHz and 1.414\,GHz, PSR J1744$-$1134 at 0.61\,GHz, PSR J1824$-$2452 at 0.61\,GHz, and PSR J1911$-$1114 at 0.41\,GHz \citep{stairs99}. The data for each pulsar occupy a row in landscape mode. From left to right, the columns show: (1) $I/I_\text{max}$ (solid curve) and $L/I_\text{max}$ (dashed curve) versus time (in s), (2) the $I$-$Q$ phase portrait, (3) the $I$-$U$ phase portrait, (4) the $Q$-$U$ phase portrait, (5) the PA swing versus time (in s) (data points with $L \geq 0.1 L_\text{max}$ and $I \geq 0.1 I_\text{max}$ plotted only). Data are presented courtesy of the EPN online archive.}
\label{data2}
\end{figure*}

\begin{figure*}
\includegraphics[scale=0.8]{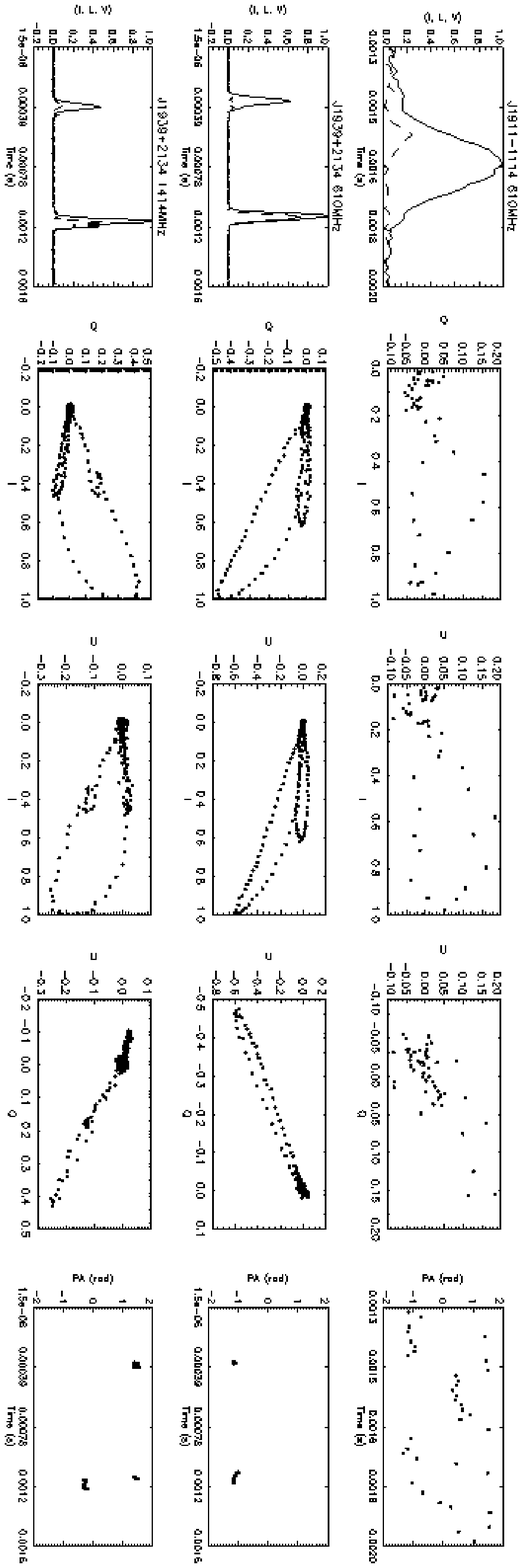}
\caption{Pulse profiles, Stokes phase portraits and PA swings for PSR J1911$-$1114 at 0.61\,GHz, and PSR J1939+2134 at 0.61\,GHz and 1.414\,GHz \citep{stairs99}. The data for each pulsar occupy a row in landscape mode. From left to right, the columns show: (1) $I/I_\text{max}$ (solid curve) and $L/I_\text{max}$ (dashed curve) versus time (in s), (2) the $I$-$Q$ phase portrait, (3) the $I$-$U$ phase portrait, (4) the $Q$-$U$ phase portrait, (5) the PA swing  versus time (in s) (data points with $L \geq 0.1 L_\text{max}$ and $I \geq 0.1 I_\text{max}$ plotted only). Data are presented courtesy of the EPN online archive.}
\label{data3}
\end{figure*}

The data in Figures \ref{data1}--\ref{data3} are too low in quality to allow detailed fits for the angles $\alpha$ and $i$ and the magnetic geometry, except for PSR J1939+2134 and PSR J0437$-$4715, which we model in detail in Sections \ref{sec:j1939} and \ref{sec:j0437}. It is still instructive, however, to compare the observed Stokes phase portraits in Figures \ref{data1}--\ref{data3} with the atlas of look-up tables  in the Appendix and make some general remarks.

With the exception of PSR J1022+1001, which has an S-shaped PA swing, the PA swings of the five other MSPs are flat or noisy, ruling out a purely dipolar magnetic geometry. Below, we list the MSPs where we have been able to find approximately matching orientations, beam patterns and linear polarization models for a current-modified dipole. The relevant look-up tables are appended in parentheses. In inferring the orientations, we follow some rules of thumb. (i) If there is an interpulse present, we limit the look-up range to $\alpha \approx i \gtrsim 60^\circ$. (ii) If there is more than one peak, we model the emission as a hollow cone. (iii) If $L$ peaks with $I$, we assume $L = I \cos \theta$, whereas, if $L$ vanishes at the pulse centroid, we assume $L = I \sin \theta$. Away from the $\alpha = i$ diagonal, both linear polarization patterns yield similar phase portraits.
\begin{enumerate}
\item PSR J1012+5307 (0.61\,GHz, Figure \ref{data1}, third row): hollow cone,  $L = I \cos \theta $, $(\alpha, i) = (70^\circ, 10^\circ)$ (Figures \ref{m25w10lcostheta_tqvsi}--\ref{m25w10lcostheta_tpa}). This object has an interpulse. The balloons in the $I$-$Q$, $I$-$U$, and $Q$-$U$ planes match approximately the orientations of the balloons in the model, although they have different widths. 
\item PSR J1713+0747 (1.414\,GHz, Figure \ref{data2}, third row): filled core beam, $L = I \sin \theta$, $(\alpha, i) = (80^\circ, 30^\circ)$ (Figures \ref{m0w10lcostheta_tqvsi}--\ref{m0w10lcostheta_tpa}). The straight lines in all three phase portraits match the model, although the gradient of the $I$-$Q$ line is less steep than in the model. 
\item J1744$-$1134 (0.61\,GHz, Figure \ref{data2}, fourth row): filled core beam, $L = I \cos \theta$, $(\alpha, i) = (30^\circ, 20^\circ)$ (Figures \ref{m0w10lcostheta_tqvsi}--\ref{m0w10lcostheta_tpa}). The straight line in the $I$-$Q$ plane matches a thin balloon in the model, while the balloons in the $I$-$U$ and $Q$-$U$ planes match balloons in the model. Note that the balloons in the model are tilted upwards ($dU/dQ > 0$), whereas in the data they are tilted downwards ($dU/dQ < 0$).
\item J1824$-$2452 (0.61\,GHz, Figure \ref{data2}, fifth row): hollow cone beam, $L = I \cos \theta$, $(\alpha, i) = (60^\circ, 60^\circ)$ (Figures \ref{m25w10lcostheta_tqvsi}--\ref{m25w10lcostheta_tpa}). The pulse profile has three peaks, suggesting that this object may have a double-peaked interpulse. The data matches the model if the $Q$-$U$ pattern is rotated by $\approx 180^\circ$.
\end{enumerate}

Where multi-frequency observations are available, we only analyse the frequency at which the phase portraits are resolved best. We emphasize that the matches are approximate, and that the figures in the Appendix show only the phase portraits at one altitude, viz. $r = 0.1 r_\text{LC}$. More detailed modelling of $I$ and $L$ as a function of emission altitude and ($\theta, \phi$) must be done to obtain more accurate matches, including the possibility that the emission originates from several altitudes \citep{johnston08}.

For PSR J1022+1001 (Figure \ref{data1}, rows 4--6), whose S-shaped PA swing is nominally dipolar, the heart shape in the $Q$-$U$ plane roughly matches a pure dipole at $(\alpha, i) \approx (70^\circ, 20^\circ)$ for a hollow cone with either polarization model. However, the observed heart shape differs slightly from the model, and $L(t)$ is actually triple-peaked, not double-peaked. Further information on the polarization basis (e.g. at several frequencies) is required in order to accurately determine the orientation and magnetic geometry. 

We now test whether the published $\alpha$ and $i$ values in Table \ref{tab:pop}, inferred from the rotating vector model, are consistent with the observed Stokes phase portraits. We refer the reader to the look-up tables in Figures \ref{m0w10lcostheta_tqvsi}--\ref{m25w10lsintheta_tuvsq} in the Appendix. For PSR J1022+1001 (Figure \ref{data1}, rows 5--6), the PA swings at 0.61\,GHz and 1.414\,GHz imply two very different orientations, namely, $(\alpha, i) = (140^\circ, 135^\circ)$ and $(83^\circ, 76^\circ)$ respectively \citep{stairs99}. Already, this is worrying, as the orientation of a given pulsar should be  unique, no matter what altitude the emission comes from. Moreover, neither of these orientations yield Stokes phase portraits which match the data, for any beam or linear polarization pattern. One can verify this easily by examining the phase portraits in the vicinity of $(\alpha, i) = (40^\circ, 50^\circ)$ and $(80^\circ, 70^\circ)$ in Figures \ref{m0w10lcostheta_tqvsi}--\ref{m25w10lsintheta_tuvsq} in the Appendix. For example, for a hollow cone with $L = I \sin \theta$, at $(80^\circ, 70^\circ)$, there are three interlocking ovals in $Q$-$U$, unlike the heart shape in Figure \ref{data1}. For PSR J1824$-$2452 (Figure \ref{data2}, fifth row), the rotating vector model predicts $(\alpha, i) = (41^\circ, 81^\circ)$. For a hollow cone, the look-up tables in Figure \ref{m25w10lcostheta_tqvsi}--\ref{m25w10lcostheta_tuvsq} show interlocking ovals in the $Q$-$U$ plane, whereas the data reveal an oval joined to a straight line. Significantly, all these discrepancies are in the shape, not the orientation of the $Q$-$U$ portrait, which is basis-independent (CM10). We comment on the published ($\alpha, i$) fits for J0437$-$4715 in Section \ref{sec:j0437}, where we model the object in detail.

\subsection{Emission altitude}
We now discuss how the pulse profiles and Stokes phase portraits evolve with frequency for PSR J1022+1001 and PSR J1939+2134, the only EPN MSPs with adequate multi-frequency data. 

At 0.41\,GHz, PSR J1022+1001 has a double-peaked intensity profile. The first intensity peak is itself double-peaked in terms of its linear polarization, resulting in three $L$ peaks overall. In the phase portraits, the first intensity peak corresponds to the bottom loop of the figure-eight in the $I$-$Q$ plane and the large balloon in the $I$-$U$ plane. At 0.61\,GHz, the second pulse is stronger than the first. In the $I$-$U$ plane, the second pulse corresponds to the long, straight tail emerging from the bottom of the balloon. 

At 0.61\,GHz, PSR J1939+2134  displays a single-peaked interpulse which peaks at $\approx 0.6 I_\text{max}$. The main pulse is also single peaked. The two pulses trace out qualitatively similar patterns on the $I$-$Q$ and $I$-$U$ planes, namely elongated balloons, whose major axes are tilted by $\approx 0^\circ$ and $\approx 20^\circ$  relative to the $Q=0$ and $U = 0$ axes respectively. At 1.414\,GHz, the peak of the interpulse drops to $\approx 0.4 I_\text{max}$, and the main pulse is double-peaked. There is a dramatic difference in $Q$ for the main pulse: the slope of the major axis of the balloon changes sign, from $dQ/dI < 0$ (0.61\,GHz) to $dQ/dI > 0$ (1.414\,GHz). The balloons of the main pulse also broaden, while those of the interpulse narrow. The second peak of the main pulse appears in the phase portraits as a kink in the $I$-$Q$ and $I$-$U$ balloons. 

For all the MSPs, the different ways in which individual peaks evolve with frequency imply that $I$ and $L$ depend on $\theta$ and $\phi$ in a complicated way. The profile components might originate from different emission regions whose magnetic geometries are different functions of $r$. As the aberration and toroidal field increase with $r$, they also distort the path $\hat{\mathbf{x}}_0(t)$, further complicating $I(t)$ and $L(t)$.

\section{A detailed example of interpulse emission: PSR J1939+2134}
\label{sec:j1939}
In this section and the next, we model the pulse and linear polarization profiles of PSR J1939+2134 and PSR J0437$-$4715 in detail and attempt to determine their orientation from their Stokes phase portraits.

We first apply the iterative recipe from CM10 to PSR J1939+2134, which has $P = 1.558$\,ms and $\dot{P} = 1.051 \times 10^{-19}$s s$^{-1}$ \citep{kaspi94}, making it the second-fastest known MSP. Data for this object, at 0.61\,GHz and 1.414\,GHz, are obtained from the EPN online archive. The data were originally published in \citet{stairs99}. We choose this object because of its strong interpulse emission. 
As mentioned in CM10, the data published in the EPN are not expressed in the canonical polarization basis described in Section \ref{sec:stokes}. Additionally, the emission altitude for this object has not been estimated at either frequency due to the flatness of its PA swing. We are therefore obliged to infer $\beta$ indirectly, from qualitative considerations, in order to bring the data of both frequencies into the canonical basis. 

To accomplish this, we make a few general observations, which provide insight into the magnetic geometry and emission pattern. Firstly, we note that the shape of the main pulse changes significantly from an asymmetric single-peaked profile at 0.61\,GHz to an asymmetric double-peaked profile at 1.414\,GHz. This suggests that the magnetic geometry and possibly the beam pattern change with emission altitude. Both pulse profiles, however, are consistent with hollow cone emission. Secondly, in the main pulse, the linear polarization follows the total intensity closely, suggesting that $L \propto \cos \theta$ is a reasonable approximation. This is also true to a lesser degree in the interpulse. 

The stellar surface of this object lies at $0.13 r_\text{LC}$, placing a lower limit on the emission altitude. If we assume arbitrarily that the data at 0.61\,GHz are already in the canonical basis, we find that, at $(\alpha, i) = (20^\circ, 80^\circ)$, the models for both pure and current-modified dipoles exhibit tilted balloons in the $I$-$U$ and $Q$-$U$ planes, similar to the data at $r \approx 0.4 r_\text{LC}$. Unfortunately, without additional information on the absolute orientation of $\mathbf{\Omega}_p$, we are limited to this assumption.
  
\begin{figure*}
\centering
\subfigure[Main pulse]
{
\label{j1939mp610}
\includegraphics[scale=0.4]{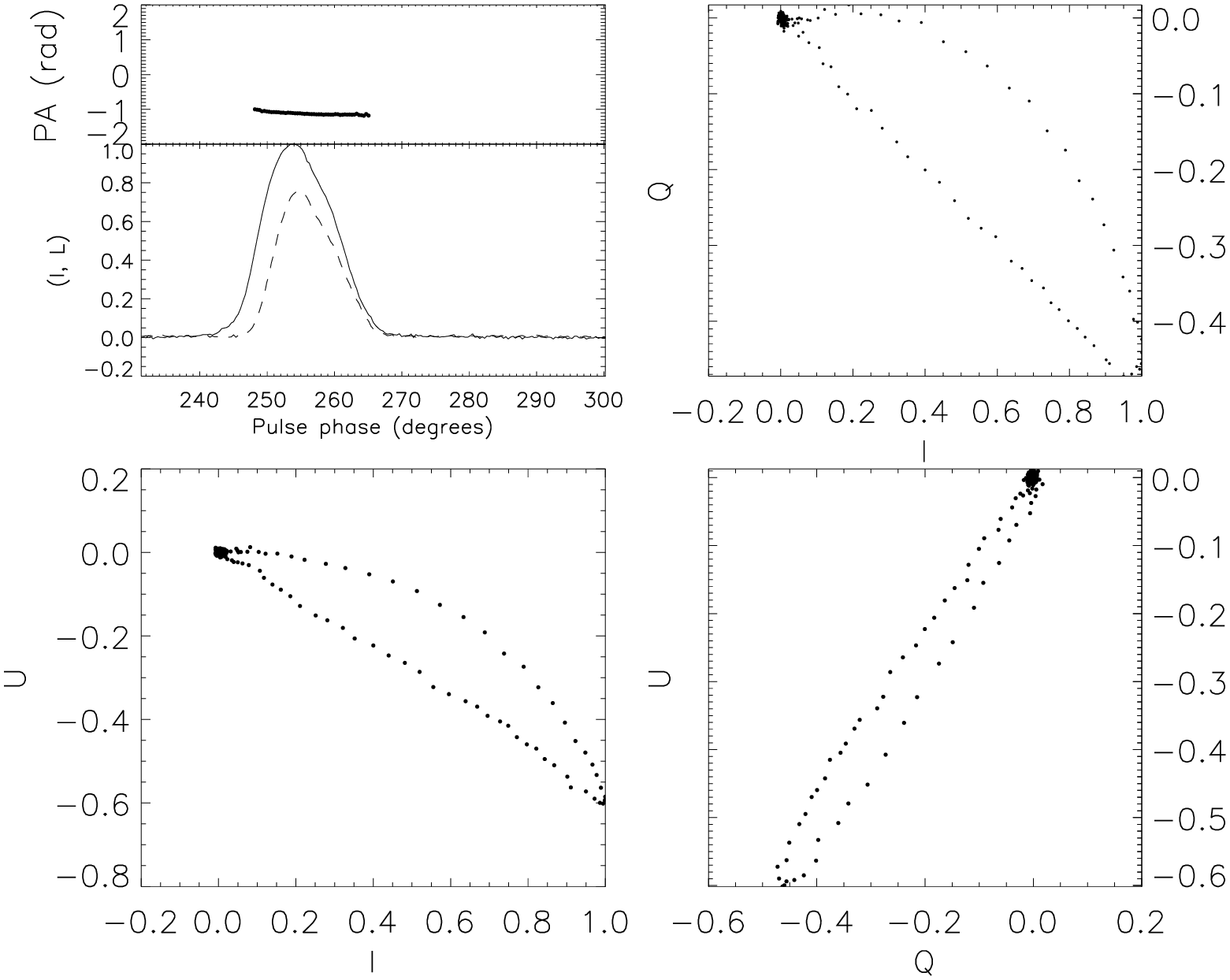}
}
\hspace{1cm}
\subfigure[Interpulse]
{
\label{j1939ip610}
\includegraphics[scale=0.4]{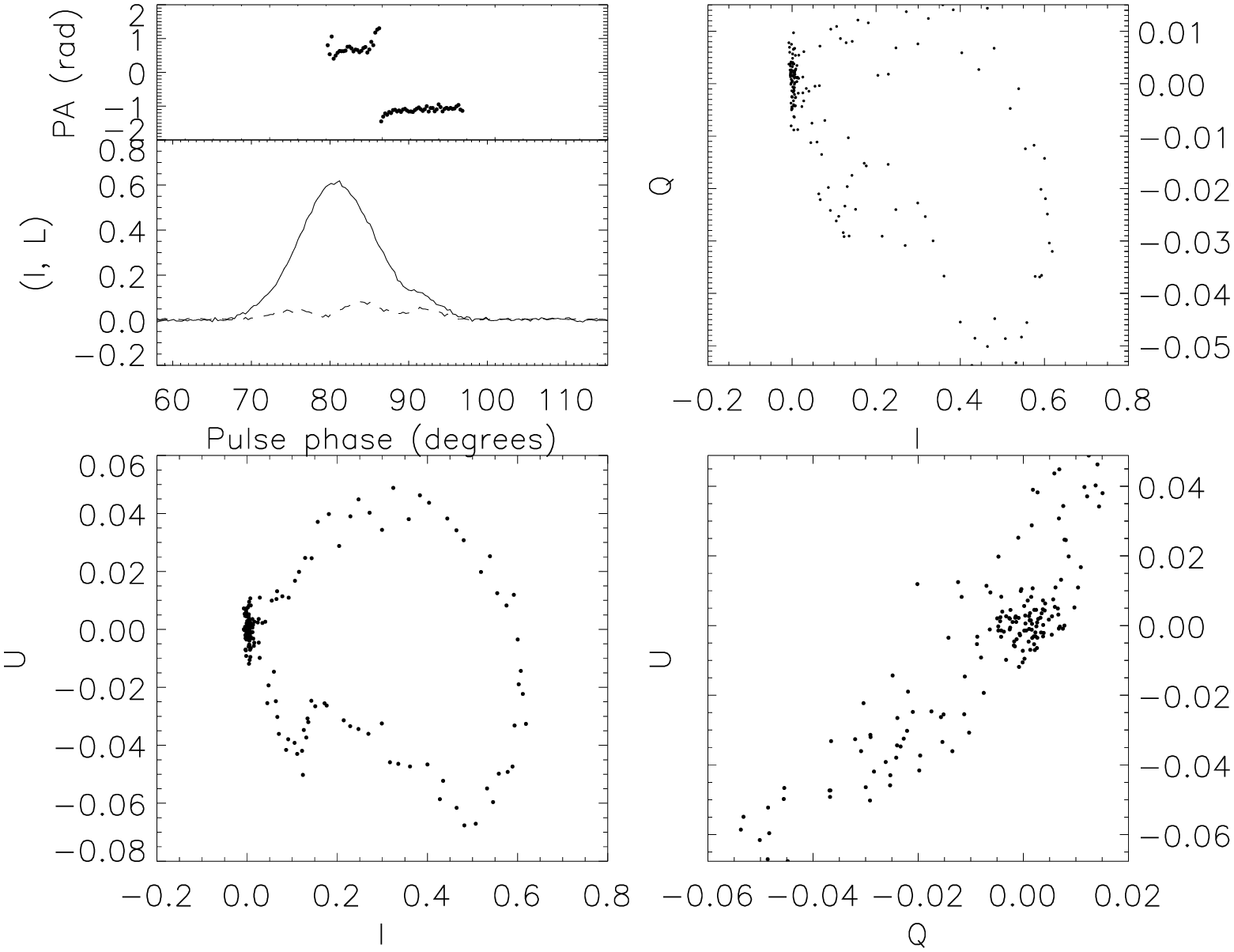}
}
\caption{Polarimetry of (a) the main pulse and (b) the interpulse of PSR J1939+2134 at 0.61\,GHz \citep{stairs99}. Each subfigure shows (clockwise from top left panel): (i) $I/I_\text{max}$ (lower subpanel, solid curve) and $L/I_\text{max}$ (lower subpanel, dashed curve) profiles and PA swing (upper subpanel, dotted curve, in rad) versus pulse phase (in degrees); (ii) $I$-$Q$ phase portrait; (iii) $Q$-$U$ phase portrait; (iv) $I$-$U$ phase portrait. Data are presented courtesy of the EPN.}
\label{j1939-610}
\end{figure*}

\begin{figure*}
\centering
\subfigure[Main pulse]
{
\label{j1939mp1414}
\includegraphics[scale=0.4]{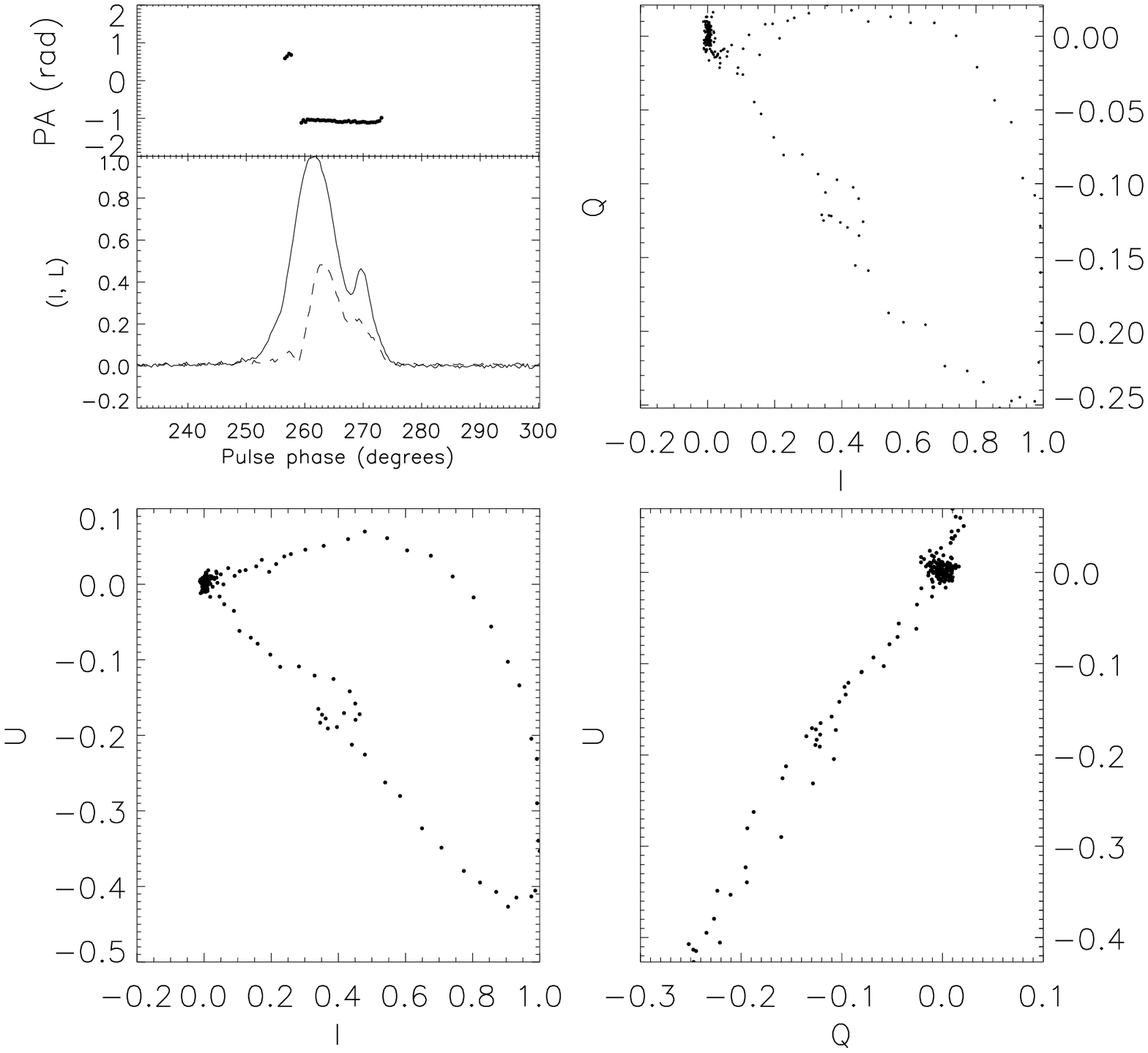}
}
\hspace{1cm}
\subfigure[Interpulse]
{
\label{j1939ip1414}
\includegraphics[scale=0.4]{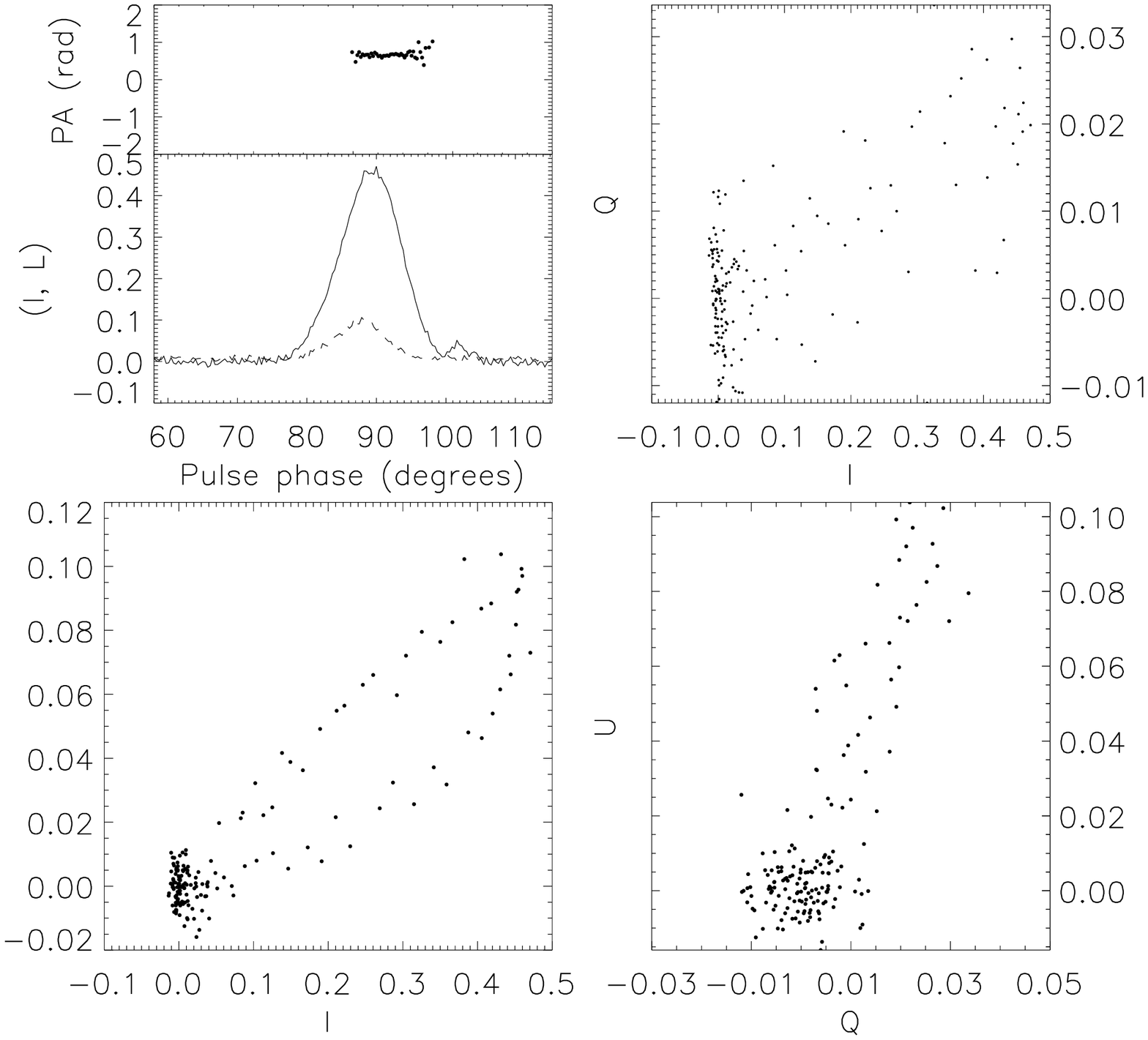}
}
\caption{As for Figure \ref{j1939-610} but at 1.414\,GHz \citep{stairs99}. Data are presented courtesy of the EPN.}
\label{j1939-1414}
\end{figure*}

Figures \ref{j1939-610} and \ref{j1939-1414} show the pulse profile, PA swing, and Stokes phase portraits of (a) the main pulse and (b) the interpulse at 0.610\,GHz and 1.414\,GHz respectively. For the 0.610\,GHz case, we assume $\beta = 0$, whereas for the 1.414\,GHz case, we align the narrow balloon shape in the $Q$-$U$ plane of Figure \ref{j1939-1414}(a) with that of Figure \ref{j1939-610}(a) by assuming $\beta = 45^\circ$. In the top left panel of each subfigure, we plot $I/I_\text{max}$ (solid curve), $L/I_\text{max}$ (dashed curve) and the PA swing (dotted curve) wherever $L \geq 0.1 L_\text{max}$. Stepping clockwise, the next three panels show $I$-$Q$, $Q$-$U$ and $I$-$U$.
 
We now examine the magnetic geometry, beam pattern, orientation, and emission altitude in more depth in Sections \ref{sec:1939magnetic}--\ref{sec:1939altitude}. We find that the model with a hollow cone and $L \propto \cos \theta$ must be generalized by letting $I$ and $L$ vary with $\phi$ in order to fit the data in detail.

\subsection{Magnetic geometry}
\label{sec:1939magnetic}
At 0.61\,GHz, the Stokes phase portraits for the main pulse are all narrow balloons. The major axes of the balloons tilt in different directions: we find $dQ/dI < 0$ in the $I$-$Q$ plane, $dU/dI < 0$ in the $I$-$U$ plane, and $dU/dQ > 0$ in the $Q$-$U$ plane. Assuming $\beta = 0$, the tilt of the $Q$-$U$ phase portrait discounts a pure dipole magnetosphere at a low emission altitude. At higher altitudes, where aberration is important, e.g. at $r = 0.1 r_\text{LC}$, the pure and current-modified dipoles produce phase portraits that are similar enough at some orientations to warrant considering both cases. 
%At 1.414\,GHz, the balloon in $I$-$Q$ broadens and changes direction so that $dQ/dI > 0$. The balloon in $I$-$U$ rotates counter-clockwise but still has $dU/dI < 0$. In $Q$-$U$, the balloon narrows to form a straight line. 

We now seek a match from the look-up tables for a hollow cone with $L = I \cos \theta$ (Figures \ref{m25w10lcostheta_tqvsi}--\ref{m25w10lcostheta_tuvsq}), keeping in mind that we are interested in orientations which provide an interpulse ($\alpha$ or $i \gtrsim 70^\circ$). We find that, at $(\alpha, i) = (20^\circ, 80^\circ)$, the phase portraits for $I$-$Q$ and $I$-$U$ match approximately the balloons in the data, although the sign of $U$ is reversed (this choice of orientation is justified in Section \ref{sec:1939orientation}). 

\subsection{Orientation $(\alpha, i)$}
\label{sec:1939orientation}
Finding the exact orientation is an iterative process, requiring the beam and polarization patterns to be adjusted at each step. Initially, we seek a match to the data at 0.61\,GHz (Figure \ref{j1939-610}). The interpulse is extremely useful in narrowing the range of possible orientations to $70^\circ \lesssim \alpha, i \lesssim 90^\circ$ (assuming beams of intrinsically equal luminosity). As the interpulse is weaker than the main pulse, we know that $\alpha$ and $i$ are less than $90^\circ$. From Figures \ref{m25w10lcostheta_tqvsi}--\ref{m25w10lcostheta_tuvsq}, there are two orientations with similar balloons in all three phase portraits, namely $(\alpha, i) = (20^\circ, 80^\circ)$. The phase portraits for a pure dipole at $(\alpha, i) = (20^\circ, 80^\circ)$ are also similar.

Before zooming in to refine the grid around $(\alpha, i) = (20^\circ, 80^\circ)$, we experiment with various emission altitudes while tailoring the pattern to fit the data. In Section \ref{sec:1939beam}, we construct beam and linear polarization patterns at $r = 0.4 r_\text{LC}$ for the data at 0.61\,GHz.

\subsection{Beam and polarization patterns}
\label{sec:1939beam}
Both the main pulse and interpulse at 0.61\,GHz are single-peaked and skewed to the left. The interpulse peaks at $\approx 0.6 I_\text{max}$. To capture this behaviour, we model the emission region as two hollow cones whose brightness varies longitudinally, i.e. the cones are shaped like horseshoes in cross-section. 
The best-fit beam pattern is given empirically by
\begin{eqnarray}
\nonumber I(\theta, \phi) &=& (2 \pi \sigma_1)^{-1/2} \left[0.8 + \lvert \sin (\phi - 1.05) \rvert \right]\\
&& \times \text{exp}\left[-0.5 (\theta - \rho_1)^2/\sigma_1^2\right]\\
\nonumber && + 0.09 (2 \pi \sigma_2)^{-1/2} \left[ 0.4 + \lvert \sin (\phi - 1.65) \rvert \right]\\
\label{eq:1939beam} && \times \text{exp}\left[-0.5 (\theta - \pi + \rho_2)^2/\sigma_2^2\right] ,
\end{eqnarray}
where $\sigma_1 = 3^\circ$ and $\sigma_2 = 3.5^\circ$ are the widths of the main pulse and interpulse respectively, and $\rho_1 = 23^\circ$ and $\rho_2 = 35^\circ$ are the corresponding opening angles. The modelled pulses are $\approx$ 3 times wider than the data as a result of the computationally limited resolution of our numerical grid.

Asymmetric emission regions are consistent with the patchy beam model introduced to explain asymmetric pulse profiles \citep{lyne88} and with theoretical models of pulsar magnetospheres like the slot gap \citep{arons83}. There are several successful precedents for pulse models with horseshoe beams, e.g. the empirical models proposed by \citet{karas07}. 

The linear polarization profiles of the main pulse and interpulse at 0.61\,GHz look surprisingly different, naively suggesting a north-south asymmetry. In the main pulse, $L$ follows the pulse profile closely, lagging the pulse centroid in phase by $\approx 4.5^\circ$, peaking at $\approx 0.8 I_\text{max}$. In the interpulse, $L$ is extremely low, peaking at $\approx 0.1 I_\text{max}$, and appears to be triple-peaked. Despite the apparent difference in the profiles, we are able to reproduce them surprisingly well using the same model, given by
\begin{eqnarray}
\label{eq:1939lprofile}
L(\theta, \phi) = \lvert \cos \theta \sin (\phi + 0.92) \rvert
\end{eqnarray}
without invoking a north-south asymmetry. As expected, however, (\ref{eq:1939lprofile}) reproduces the $L$ profile of the main pulse more accurately than that of the interpulse.
We emphasize that (\ref{eq:1939beam}) and (\ref{eq:1939lprofile}) are certainly not unique and do not fit the data exactly, but they are adequate for the empirical task at hand.

Adopting (\ref{eq:1939beam}) and (\ref{eq:1939lprofile}), we generate zoomed-in look-up tables for both pure and current-modified dipoles, in the range $14^\circ \leq \alpha \leq 24^\circ$, $76^\circ \leq i \leq 84^\circ$, with a resolution of $2^\circ$. We find the closest match is for a current-modified dipole at $(\alpha, i) = (22^\circ, 80^\circ)$, with a `by eye' uncertainty of $\pm 2^\circ$ for $\alpha$ and $\pm 1^\circ$ for $i$. This margin would widen if $I(\theta, \phi)$ and $L(\theta, \phi)$ were adjusted for each orientation.

\begin{figure*}
\centering
\subfigure[Main pulse]
{
\label{1939match610mp}
\includegraphics[scale=0.35]{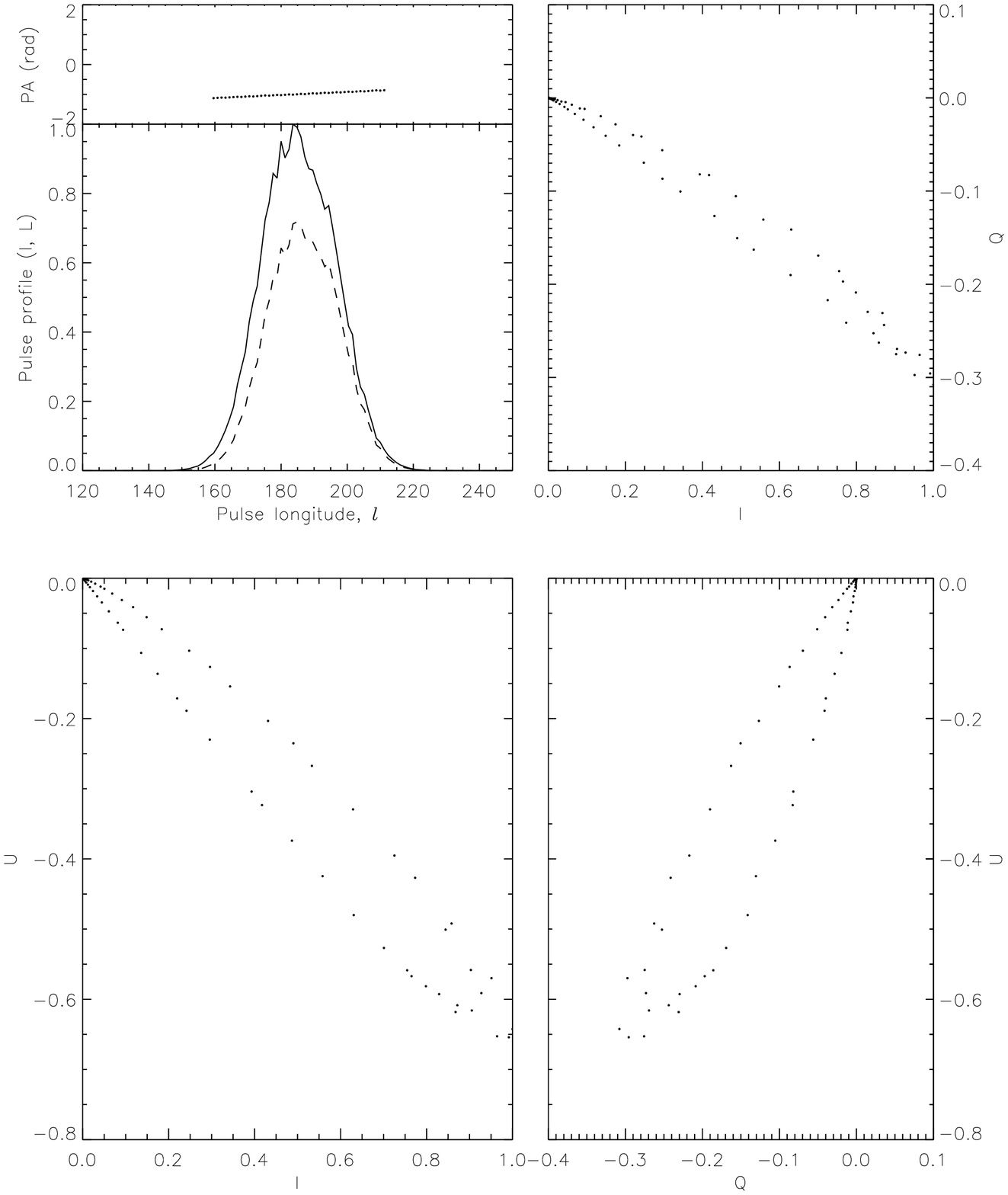}
}
\hspace{1cm}
\subfigure[Interpulse]
{
\label{1939match610ip}
\includegraphics[scale=0.35]{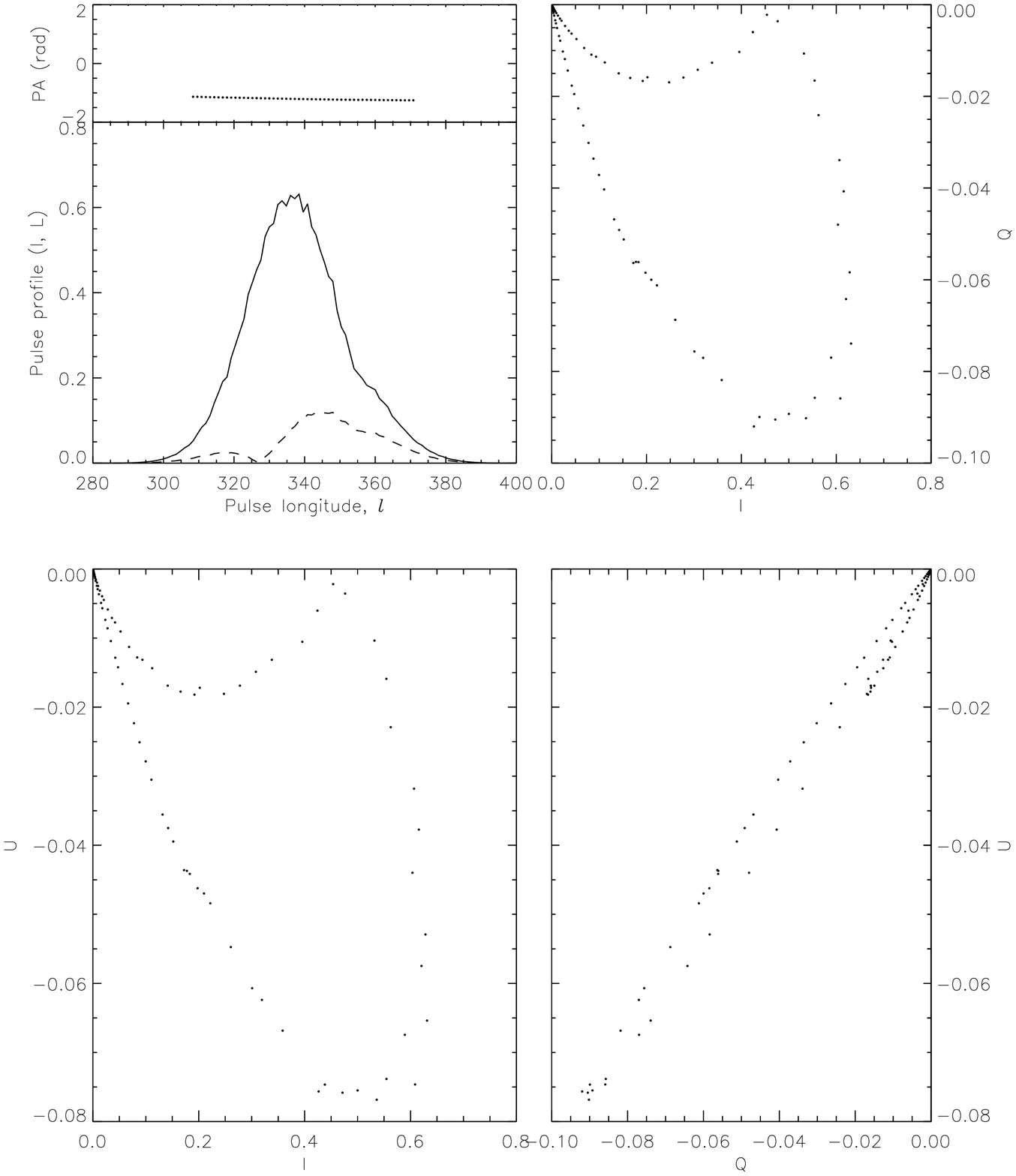}
}
\caption{Theoretical polarization model of (a) the main pulse and (b) the interpulse of PSR J1939+2134 for a current-modified dipole emitting at $r = 0.4 r_\text{LC}$, with orientation $(\alpha, i) = (22^\circ, 80^\circ)$, beam pattern given by (\ref{eq:1939beam}), and linear polarization given by (\ref{eq:1939lprofile}). Each subfigure shows (clockwise from top left panel): (i) $I/I_\text{max}$ (lower subpanel, solid curve) and $L/I_\text{max}$ (lower subpanel, dashed curve) profiles and PA swing (upper subpanel, dotted curve, in rad) versus pulse phase (in degrees); (ii) $I$-$Q$ phase portrait; (iii) $Q$-$U$ phase portrait; (iv) $I$-$U$ phase portrait. }
\label{1939match610}
\end{figure*}

In Figure \ref{1939match610}, we plot the pulse profile, PA swing and Stokes phase portraits of the model at $r = 0.4 r_\text{LC}$ and $(\alpha, i) = (22^\circ, 80^\circ)$. The slight jaggedness of the pulse profiles is a product of the finite grid resolution. The Stokes phase portraits of the main pulse [Figure \ref{1939match610mp}] match the data in Figure \ref{j1939mp610} reasonably well. In the data, the $I$-$Q$ balloon ranges from $-0.5 \lesssim Q \lesssim 0$, whereas in the model it is thinner and ranges from $-0.3 \lesssim Q \lesssim 0$. The $I$-$U$ balloon in the data ranges from $-0.6 \lesssim U \lesssim 0$, whereas in the model it ranges from $-0.65 \lesssim U \lesssim 0$. The PA swing in the data is nearly flat, with a slight negative gradient, whereas the model shows a slight positive gradient.
For the interpulse, there is poorer agreement in $L$. The tilted balloon in $I$-$Q$ from the data [Figure \ref{j1939ip610}] is reproduced in Figure \ref{1939match610ip}, including the kink visible at $(I, Q) \approx (0.2, -0.03)$. In the $I$-$U$ plane, the data feature a tilted balloon, with a prominent kink at $(I, U) \approx (0.2, -0.02)$. Our model shows a broad hockey stick instead. At a stretch, one may perhaps argue that the balloon with the kink resembles the hockey stick qualitatively, but we do not press the point. We note that if the upper half of the hockey stick is reflected about $U = 0$, it would match more closely. In the $Q$-$U$ plane, the tilted oval seen in the data is reproduced in the model, but with $-0.07 \lesssim U \lesssim 0.05$ in the data, and $-0.08 \lesssim U \lesssim 0$ in the model. The modelled PA swing is flat, with a negative gradient, and lacks the phase-wrapping seen in the data.

\subsection{Emission altitude}
\label{sec:1939altitude}

\begin{figure*}
\includegraphics[scale=0.7]{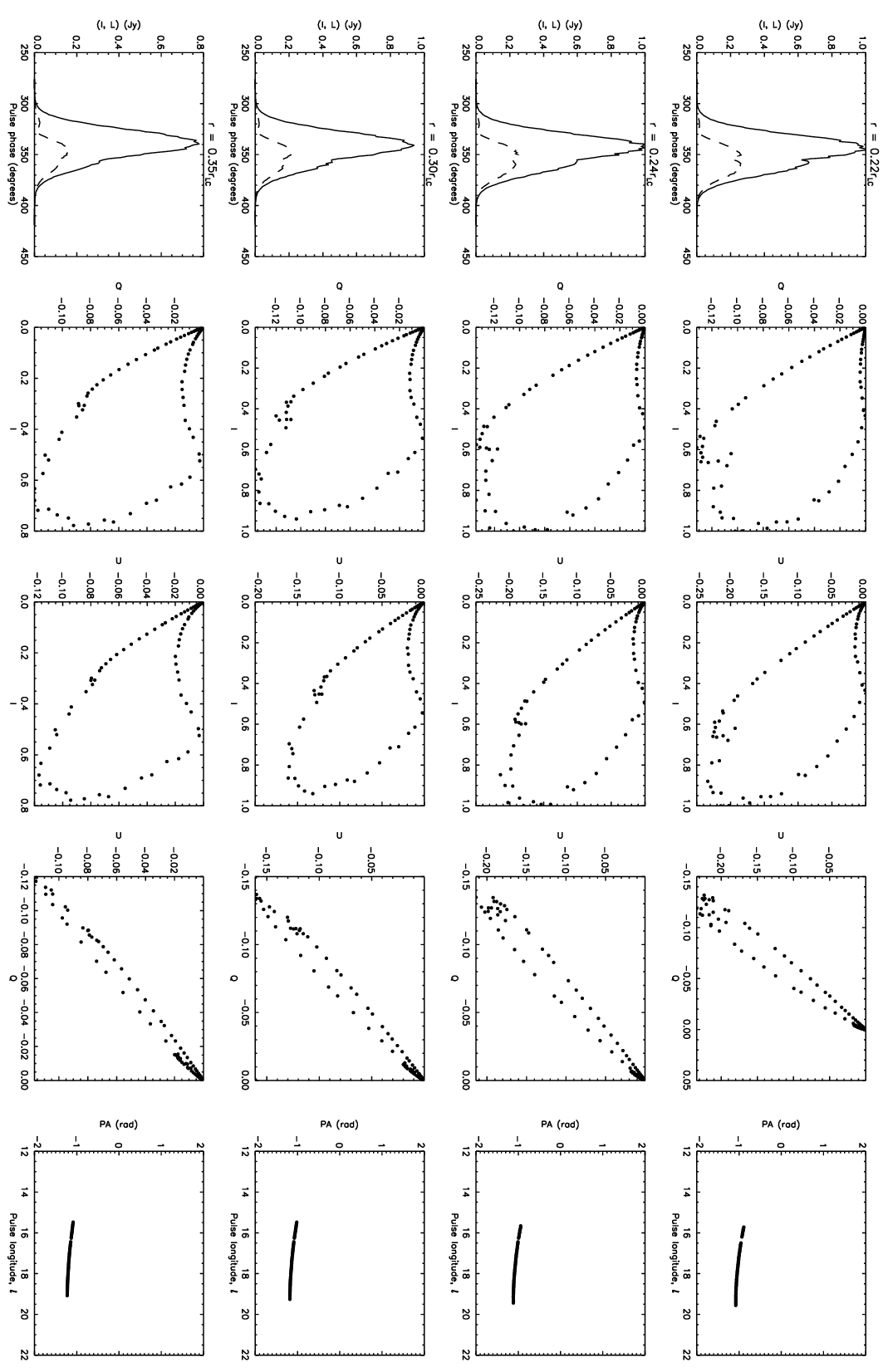}
\caption{Theoretical polarization model of Pulse 1 of PSR J1939+2134 as a function of emission altitude, for a current-modified dipole with $(\alpha, i) = (22^\circ, 80^\circ)$. In landscape mode, the plots for each emission altitude occupy rows, increasing from $r = 0.22 r_\text{LC}$ (top row) to $r = 0.35 r_\text{LC}$ (bottom row). From left to right, the columns show (1) $I/I_\text{max}$ (solid curve) and $L/I_\text{max}$ (dashed curve) versus pulse phase $l$ (in units of degrees), (2) $I$-$Q$ phase portrait, (3) $I$-$U$ phase portrait, (4) $Q$-$U$ phase portrait, and (5) the PA swing (in rad; data points with $L \geq 0.1 L_\text{max}$ plotted only).}
\label{1939-allmp}
\end{figure*}

\begin{figure*}
\includegraphics[scale=0.7]{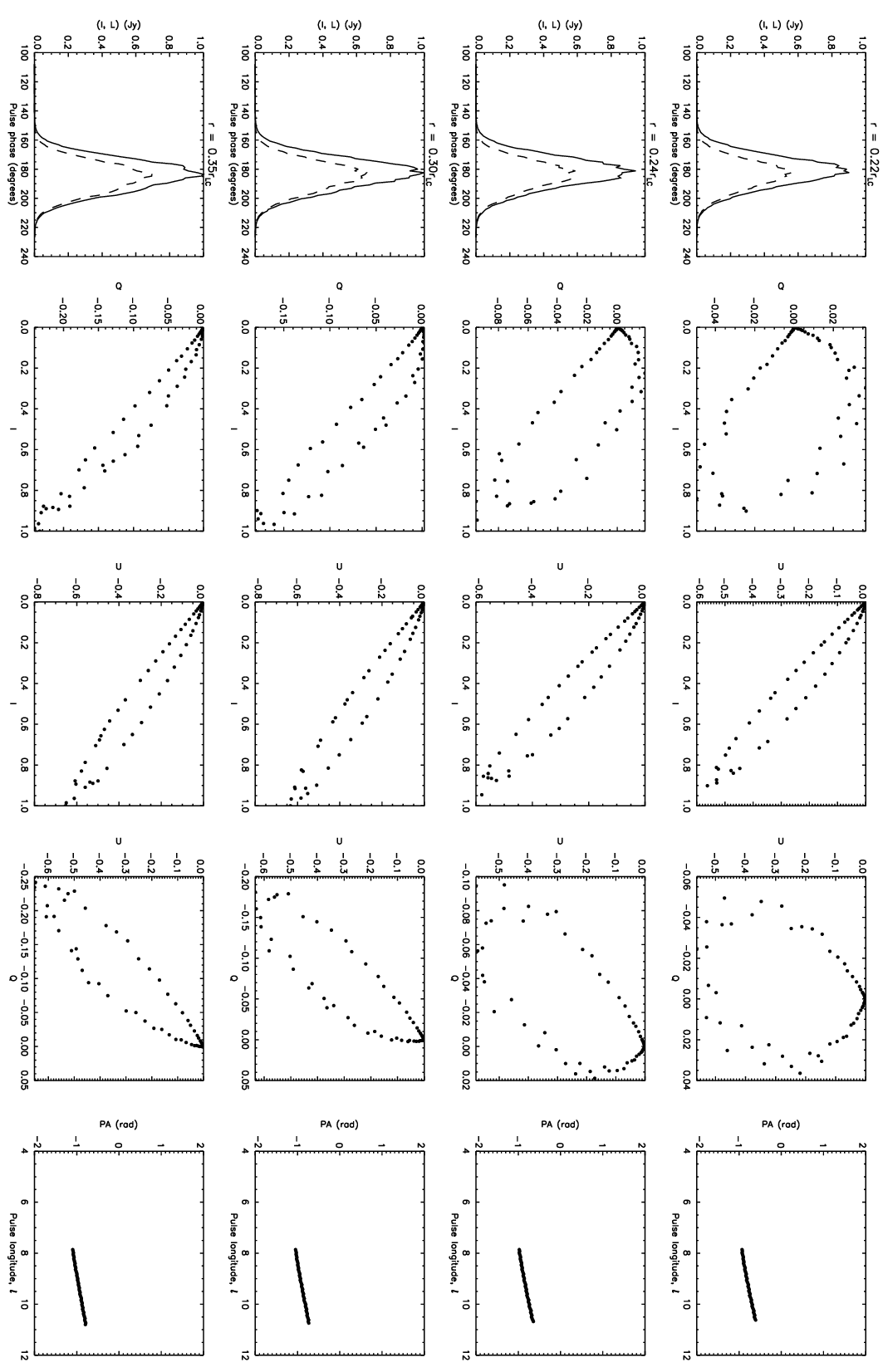}
\caption{As for Figure \ref{1939-allmp} but for Pulse 2 of PSR J1939+2134.} 
\label{1939-allip}
\end{figure*}

According to the standard radius-to-frequency mapping, the observation frequency scales with emission radius as $r^{-3/2}$ \citep{ruderman75, cordes78}. If the data at 0.61\,GHz correspond to $r = 0.4 r_\text{LC}$, then 1.414\,GHz corresponds to $r = 0.22 r_\text{LC}$. Figures \ref{1939-allmp}--\ref{1939-allip} show the pulse profiles, PA swing and Stokes phase portraits predicted theoretically for both pulses, for emission altitudes ranging from $r = 0.22r_\text{LC}$ to $0.35r_\text{LC}$. The relative heights of the pulses change with emission altitude. We label them Pulse 1 (Figure \ref{1939-allmp}), corresponding to the main pulse in the data, and Pulse 2 (Figure \ref{1939-allip}), corresponding to the interpulse in the data.

The theoretical pulse profile and phase portraits at $r = 0.22 r_\text{LC}$ (Figures \ref{1939-allmp}--\ref{1939-allip}, top row) display some interesting features. First, the main pulse and interpulse have roughly the correct shapes, but swap positions in phase, i.e. the hollow cone which emits the main pulse at $r = 0.4 r_\text{LC}$ also emits the interpulse at $r = 0.22 r_\text{LC}$, and vice versa. Upon inspection, it is likely that the same is true in the data. The triple-peaked linear polarization profile seen in Figure \ref{j1939ip610} is also present in Figure \ref{j1939mp1414}, although the first component in $L$ is much weaker than the second and third at 1.414\,GHz. Additionally, the kinks seen in the $I$-$Q$ and $I$-$U$ planes of Figure \ref{j1939ip610} are seen in Figure \ref{j1939mp1414}.
In the data, the interpulse peaks at $\approx 0.5 I_\text{max}$, compared to $\approx 0.9 I_\text{max}$ in the model at $r = 0.22 r_\text{LC}$.

Second, the linear polarization profile and phase portraits at $r = 0.22 r_\text{LC}$ reproduce the main pulse reasonably well but do not match the interpulse.
 The linear polarization of the main pulse (Pulse 1; Figure \ref{1939-allmp}, top row) is $\approx 50$\% weaker than observed [Figure \ref{j1939mp1414}]. Also, in the simulated profile, the third $L$ peak in the main pulse is comparable in height to the second peak, while in the data it is weaker. In the simulated $I$-$Q$ and $I$-$U$ planes of the main pulse, there are reasonable matches to the balloons in the data. In the simulated $I$-$Q$ plane, we see a balloon with a kink at $(I, Q) \approx (0.6, -0.1)$. In the data, the kink appears at $(I, Q) \approx (0.4, -0.1)$. In the $I$-$U$ plane, the kink seen in the data at $(I, U) \approx (0.4, -0.1)$ is reproduced at $(I, U) \approx (0.6, -0.2)$ in the model. In the $Q$-$U$ plane, the data trace out a thin balloon with $dU/dQ > 0$ spanning $-0.3 \lesssim Q \lesssim 0$ and $-0.4 \lesssim U \lesssim 0.05$. The simulated phase portrait shows a thin balloon with the same orientation, spanning $-0.15 \lesssim Q \lesssim 0$ and $-0.3 \lesssim U \lesssim 0$.
For the interpulse (Pulse 2; Figure \ref{1939-allip}, top row) the simulated total intensity is twice the observed intensity, and the simulated linear polarization is $\approx$ 6 times stronger than observed [Figure \ref{j1939ip1414}]. The simulated and observed balloons in the $I$-$Q$ and $I$-$U$ planes are rotated by $90^\circ$ clockwise with respect to the data, whereas the $Q$-$U$ balloon is rotated by $180^\circ$. These discrepancies are also reflected in the PA swing.

As the emission altitude increases from $r = 0.22 r_\text{LC}$ to $r = 0.35 r_\text{LC}$, the phase portraits of the main pulse (Pulse 1; Figure \ref{1939-allmp}) change. In the $I$-$Q$ plane, the kink in the balloon shifts towards $(I, Q) \approx (0.3, -0.08)$. At $r = 0.35r_\text{LC}$, the balloon in the $I$-$U$ plane starts to resemble the hockey stick seen in Figure \ref{1939match610ip}. In $Q$-$U$, the thin balloon rotates clockwise. 
In the interpulse (Pulse 2; Figure \ref{1939-allip}), the $I$-$Q$ balloon narrows and lengthens in $Q$, and the $Q$-$U$ balloon narrows.

From Figures \ref{1939-allmp}--\ref{1939-allip}, we draw the following conclusions. (i) Although the simple model given by (\ref{eq:1939beam}) and (\ref{eq:1939lprofile}) models the 0.61\,GHz data reasonably successfully, it fails for the data at 1.414\,GHz displayed in Figure \ref{j1939-1414}. However, the observed pulse profiles and Stokes phase portraits suggest that the emission region of the main pulse at 0.61\,GHz corresponds to that of the interpulse at 1.414\,GHz, and vice versa. Additionally, the emission pattern may change with radius. (ii) The discrepancies between the data and the phase portraits at $r = 0.22r_\text{LC}$ (the altitude predicted by the radius-to-frequency mapping) indicate that the toroidal field may not increase monotonically with $r$. The phase portraits for the interpulse between $r = 0.22r_\text{LC}$ and $r = 0.35r_\text{LC}$ are all a poor match to the data. (iii) It is possible that the data should be referred to a different value of $\beta$ at 1.414\,GHz than the one we assume, which would rotate the $Q$-$U$ phase portrait, and change the shapes of the $I$-$Q$ and $I$-$U$ patterns. 


\section{A detailed multi-peaked example: PSR J0437$-$4715}
\label{sec:j0437}

We now repeat the procedure in Section \ref{sec:j1939} and CM10 for PSR J0437$-$4715. This object has $P = 5.758$\,ms and $\dot{P} = 5.729 \times 10^{-20}$\,s s$^{-1}$ \citep{bell97}. It was chosen because it exhibits five distinct peaks in its pulse profile, clearly visible at 1.44\,GHz, and a highly structured PA swing.  There is no interpulse observed in this object. 
Unlike the other objects considered in this paper and CM10, we find that PSR J0437$-$4715 cannot be modelled by either a pure or a current-modified dipole field, even if a multiple-peaked beam pattern is constructed emiprically to fit the $I(t)$ data exactly. Indeed, the Stokes phase portraits point persuasively to the existence of a strong quadrupole and higher-order multipoles at the radio emission altitude. In this respect, PSR J0437$-$4715 is an excellent candidate for more detailed Stokes tomography studies in the future. In this section, we restrict ourselves to presenting the argument that the pure and current-modified dipoles categorically fail to match the data for the polarization models that work well for the other objects studied in this paper and CM10.

\begin{figure}
\includegraphics[scale=0.4]{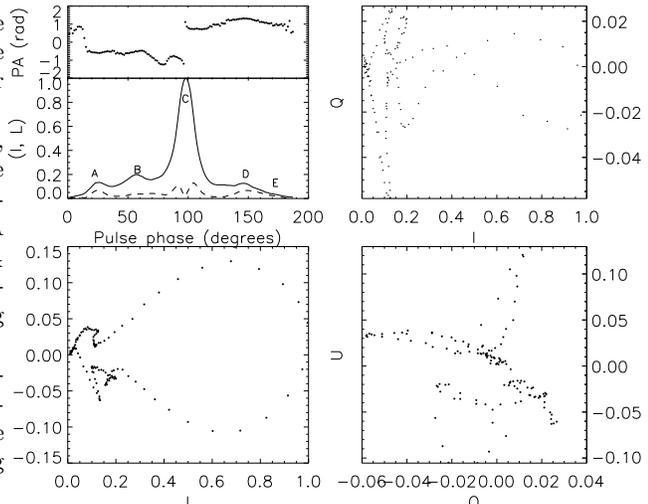}
\caption{Polarimetry of PSR J0437$-$4715 at 1.44\,GHz \citep{manchester95}. Clockwise from top left panel: (a) $I/I_\text{max}$ (lower subpanel, solid curve) and $L/I_\text{max}$ (lower subpanel, dashed curve) profiles, and PA swing (upper subpanel, dotted curve, in rad) all plotted against pulse phase (in degrees); (b) $I$-$Q$ phase portrait; (c) $Q$-$U$ phase portrait; (d) $I$-$U$ phase portrait. Data are presented courtesy of the EPN.}
\label{j0437}
\end{figure}

Data for PSR J0437$-$4715, at 1.44\,GHz and 4.6\,GHz, are obtained from the EPN online archive. The data were originally published by \citet{manchester95}.
Figure \ref{j0437} presents the pulse profile, phase portraits, and PA swing (in rad) at 1.4\,GHz. The top left panel shows $I/I_\text{max}$ (solid curve), $L/I_\text{max}$ (dashed curve) and the PA swing (dotted curve, at longitudes where $L \geq 0.1 L_\text{max}$). Stepping clockwise, the next three panels show $I$-$Q$, $I$-$U$ and $Q$-$U$. 

As with PSR 1939+2134, we assume that the data obtained from the EPN are presented in the canonical polarization basis at one reference frequency, chosen here to be 1.44\,GHz. The stellar surface is at $r = 0.036 r_\text{LC}$, providing a lower limit on the emission altitude. 

The pulse profile has five components, labelled A--E in Figure \ref{j0437}, two on either side of the largest peak (C). Each component in the pulse profile corresponds to a distinct sub-pattern in the Stokes phase portraits. In the $I$-$Q$ plane, peaks A and B correspond to the two small loops at $Q > 0$, peak C is the large figure-eight, and peaks D and E correspond to the loop at $Q < 0$. In the $I$-$U$ plane, peaks A and B correspond to the kinks at $U < 0$, peak C is the large balloon, and peaks D and E correspond to the kink at $U > 0$. The $U$-$Q$ plane is complicated, forming a rough X-shape, with one diagonal having $dU/dQ > 0$ (we call this diagonal 1), and the other having $dU/dQ < 0$ (diagonal 2). Diagonal 1 corresponds to peak C. In diagonal 2, peaks A and B occupy the $Q > 0$ region, whereas peaks D and E occupy the $Q < 0$ region.

 The linear polarization within peak C is double-peaked. This kind of structure is common and is modelled adequately by a filled core beam with $L = I \sin \theta$, as demonstrated for several objects in CM10. The phase separation of the peaks suggests that A, B, D, and E originate from two hollow cones centred on the same axis (peak C). Peak pairs B/C and C/D are separated by $\approx 0.7$ rad and $\approx 0.9$ rad respectively, while A/C and C/E are both separated by $\approx 1.2$ rad. We confirm a posteriori that $\alpha$ and $i$ lie in ranges where interpulse emission does not contribute significantly.

\subsection{Magnetic geometry}
In order to determine the magnetic geometry, we search the look-up tables for a good match involving a filled core and a hollow cone beam. We assume $L = I \sin \theta$ because of the double-peaked $L$ profile in peak C. The phase portraits for the filled core should match the large patterns corresponding to peak C, while the phase portraits for the hollow cone should match the smaller patterns. We do not expect perfect matches due to the complex beam and polarization patterns. At a minimum, however, we seek an approximate match for the rough figure-eight that forms diagonal 1 in the $Q$-$U$ plane, the figure-eight in $I$-$Q$, and the balloon in $I$-$U$.

The pure dipole look-up tables at $r = 0.1 r_\text{LC}$ do not feature figure-eight shapes in the $I$-$Q$ plane for any orientation. For $\alpha < i$, there are balloons in $I$-$U$, and heart shapes in $Q$-$U$. As in many other opbjects, a pure dipole is ruled out (CM10).

For the current-modified dipole (Figures \ref{m0w10lsintheta_tqvsi}--\ref{m0w10lsintheta_tuvsq}), the closest match is at $(\alpha, i) = (30^\circ, 30^\circ)$ (see Section \ref{sec:0437orientation} for a detailed justification). For a filled core, there are asymmetric figure-eights in the $Q$-$U$ and $I$-$U$ planes (Figures \ref{m0w10lsintheta_tqvsi}--\ref{m0w10lsintheta_tuvsq}), and a broad oval in $I$-$Q$ (Figure \ref{m0w10lsintheta_tqvsi}). The hollow cone phase portraits at this orientation feature asymmetric mosquitoes in $I$-$Q$ and $I$-$U$ (Figure \ref{m25w10lsintheta_tqvsi}--\ref{m25w10lsintheta_tuvsi}) and an asymmetric heart shape in $Q$-$U$ (Figure \ref{m25w10lsintheta_tuvsq}). At this stage, we cannot confidently discount either magnetic configuration, as the Stokes portraits change when the filled core and hollow cones are combined. This issue is examined thoroughly in Sections \ref{sec:0437beam}--\ref{sec:0437multimodal}.

\subsection{Orientation}
\label{sec:0437orientation}
In this section, we justify $(\alpha, i) = (30^\circ, 30^\circ)$ as the best matching orientation.
As PSR J0437$-$4715 does not have an observed interpulse, we rule out all orientations with $\alpha \geq 80^\circ, i \geq 80^\circ$. We also rule out orientations with $i > \alpha$ because the associated phase portraits look nothing like the data. For example, for a filled core, Figures \ref{m0w10lsintheta_tqvsi}--\ref{m0w10lsintheta_tuvsq} contain balloons in $I$-$Q$, narrow, tilted balloons and straight lines in $I$-$U$, and ovals in $U$-$Q$. None of these patterns appear in the data in Figure \ref{j0437}. The $Q$-$U$ discrepancy is especially significant as the shape of the $Q$-$U$ portrait is basis-independent.

The best matching orientation predicted by the RVM is $(\alpha, i) = (145^\circ, 140^\circ)$ \citep{manchester95}, although the authors note that the PA swing deviates largely from the model. For this reason, no formal uncertainties are assigned to the fitted parameters, which were chosen to be a reasonable representation of the data. For a dipole field, the Stokes phase portraits are symmetric about $(\alpha, i) = (90^\circ, 90^\circ)$, i.e. the phase portraits for $(\alpha, i) = (145^\circ, 140^\circ)$ and $(35^\circ, 40^\circ)$ are identical. From CM10, the phase portraits for a pure dipole at $(\alpha, i) = (40^\circ, 40^\circ)$ and $r \ll r_\text{LC}$, with a hollow cone and $L = \sin \theta$, feature a narrow, tilted balloon in $I$-$Q$, a mosquito in $I$-$U$, and a heart in $Q$-$U$ (see Figures 22--24 in CM10). The $I$-$U$ and $Q$-$U$ shapes are symmetric about $U = 0$. Interestingly, although we have chosen our best match independently of the RVM results, the two orientations are close.  

 We reiterate that the complex multiple-peaked beam and polarization patterns complicate the matching process. Some orientations must be tested with beam patterns tailored to fit the data, as described in Section \ref{sec:0437beam}, before being ruled out.
For $(\alpha, i) = (30^\circ, 30^\circ)$, the appropriate beam pattern is a filled core surrounded by two hollow cones. The resulting phase portraits show distorted, tilted balloons in both $I$-$Q$ and $I$-$U$. These shapes resemble roughly the data in Figure \ref{j0437}, although there are large discrepancies too, chiefly that the figure-eight in $I$-$Q$ is missing, and that $I$-$U$ is not symmetric about $U = 0$. In Sections \ref{sec:0437beam} and \ref{sec:0437multimodal}, we construct detailed beam and linear polarization models in an attempt to improve the fits.

\subsection{Beam pattern}
\label{sec:0437beam}
In fitting the complex pulse profile of PSR J0437$-$4715 at 1.44\,GHz, \citet{gangadhara08} identified 11 Gaussian components. They proposed that the beam pattern comprises five nested cones at different altitudes centred on the filled core, and that the altitudes range from 0.07$r_\text{LC}$--0.3$r_\text{LC}$. 

We consider a simpler model and focus on one fixed altitude. We model the pulse profile empirically with a filled core, $I_1(\theta, \phi)$ (peak C), surrounded by two hollow cones, $I_2 (\theta, \phi)$ (peaks B and D) and $I_3 (\theta, \phi)$ (peaks A and E). The intensity maps take the \textit{empirical} forms
\begin{eqnarray}
\nonumber I_1 (\theta, \phi) &=& (2 \pi \sigma_1^2)^{-1/2} \left\{ \text{exp}\left[-\theta^2/(2 \sigma_1^2)\right]  \right. \\
\label{eq:0437I1} &&\left. + \text{exp}\left[-(\theta - \pi)^2/(2 \sigma_1^2)\right] \right\},\\
\nonumber I_2 (\theta, \phi) &=& \beta_2(\phi) (2 \pi \sigma_2^2)^{-1/2} \left\{ \text{exp}\left[-(\theta - \rho_2)^2/(2 \sigma_2^2)\right]  \right. \\
\label{eq:0437I2} &&\left. + \text{exp}\left[-(\theta - \pi + \rho_2)^2/(2 \sigma_2^2)\right] \right\},\\
\nonumber I_3 (\theta, \phi) &=& \beta_3(\phi) (2 \pi \sigma_3^2)^{-1/2} \left\{ \text{exp}\left[-(\theta - \rho_3)^2/(2 \sigma_3^2)\right]  \right. \\
\label{eq:0437I3} &&\left. + \text{exp}\left[-(\theta - \pi + \rho_3)^2/(2 \sigma_3^2)\right] \right\},
\end{eqnarray}
where $\sigma_1 = 6.5^\circ, \sigma_2 = 2.5^\circ$ and $\sigma_3 = 2^\circ$ are the beam widths of the core and cones, $\rho_2 = 18^\circ$ and $\rho_3 = 26^\circ$ are the opening angles of the two cones, and $\beta_2 (\phi)$ and $\beta_3 (\phi)$ are functions describing the longitudinal structure of the two cones, given empirically by
\begin{eqnarray}
\beta_2 (\phi) &=& 0.06 \lvert \cos (0.3 \phi) \rvert,\\
\beta_3 (\phi) &=& 0.06 \lvert \cos (0.75 \phi) \rvert.
\end{eqnarray}
As in Section \ref{sec:j1939}, the cones are shaped like horseshoes.
Given (\ref{eq:0437I1})--(\ref{eq:0437I3}), we also find that the linear polarization pattern is fitted empirically by
\begin{eqnarray}
\nonumber L (\theta, \phi) &=& (3 \theta)^{-1} \sin (\theta - 0.01) + 0.07 \lvert \cos (0.75 \phi) \rvert \\
\label{eq:0437L} && \times (2 \pi \sigma_2^2)^{-1/2} \text{exp}\left[-(\theta - \rho_2)^2/(2 \sigma_2^2)\right]  .
\end{eqnarray}

We emphasize that equations (\ref{eq:0437I1})--(\ref{eq:0437L}) are not unique fits, nor do they produce perfect agreement with the data. In particular, the data show that the B/C peaks are closer to each other than C/D, yet we are unable to reproduce this with a reasonably simple model. Our modelled peaks are equidistant.  
The models are sensitive to the pulsar's orientation. Every time we vary $\alpha$ or $i$ around $(30^\circ, 30^\circ)$, we must adjust the coefficients in (\ref{eq:0437I1})--(\ref{eq:0437L}). We find that the closest match to the data, although poor, is achieved at $(\alpha, i) = (32^\circ, 26^\circ)$. We note as well that the Stokes phase portraits match marginally better if we rotate the polarization basis by $\beta = 90^\circ$. The difficulty in achieving a good match may well be telling us that the underlying magnetic geometry is not a current-modified dipole.

\subsection{Decomposed phase portraits}
\label{sec:0437multimodal}
\begin{figure*}
\includegraphics[scale=0.8]{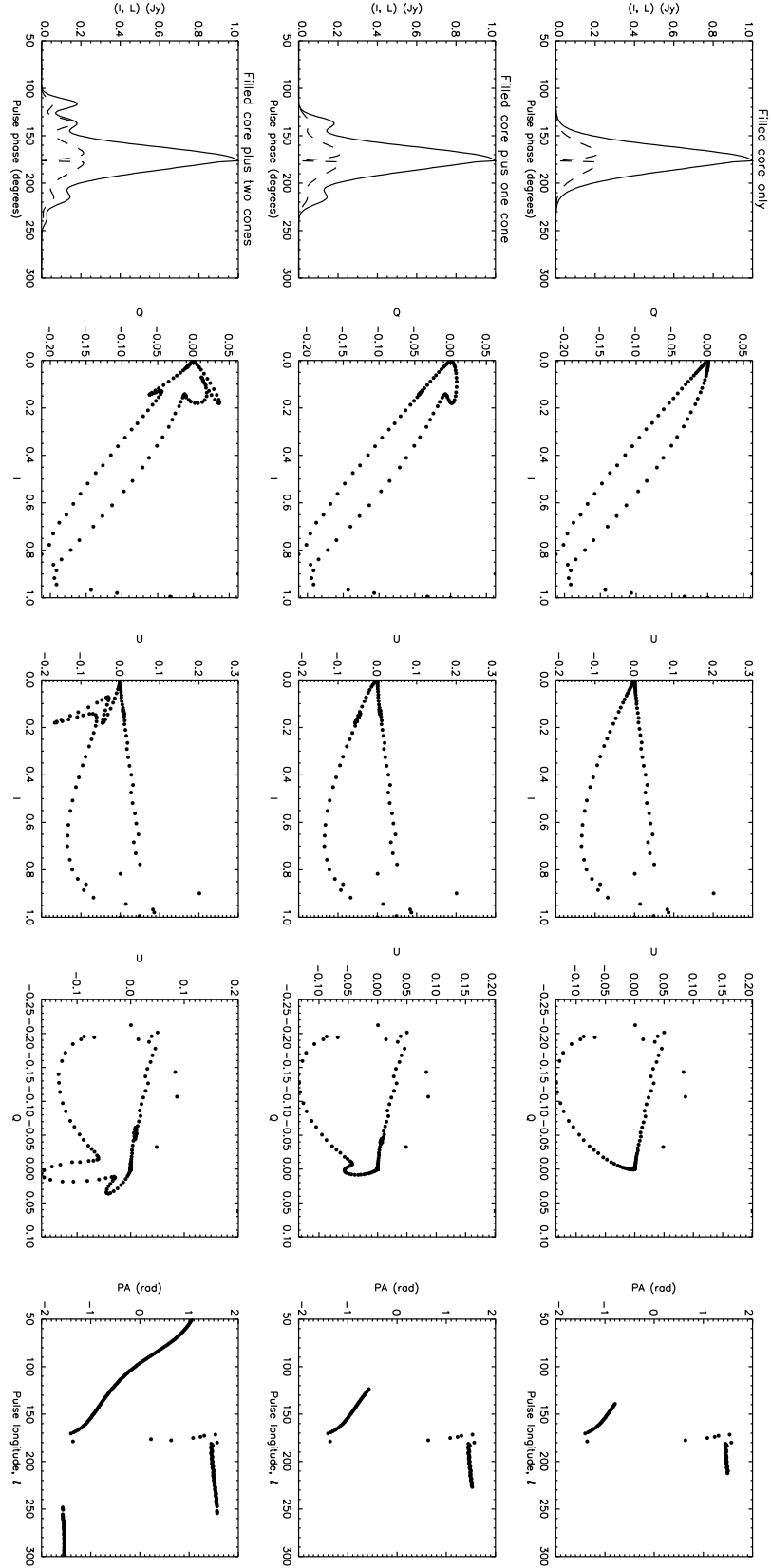}
\caption{Bottom-up, component-wise assembly of theoretical polarization models for PSR J0437$-$4715 for a current-modified dipole emitting at $r = 0.1 r_\text{LC}$ with $(\alpha, i) = (32^\circ, 26^\circ)$ and $\beta = 90^\circ$. In landscape orientation, from top to bottom, we plot the following beam patterns: (top row) peak C, filled core, equation (\ref{eq:0437I1}); (middle row) peaks B--D, filled core plus hollow cone, equations (\ref{eq:0437I1}) and (\ref{eq:0437I2}); and (bottom row) peaks A--E, filled core plus two hollow cones, equations (\ref{eq:0437I1}), (\ref{eq:0437I2}) and (\ref{eq:0437I3}). Linear polarization in all three rows is given by (\ref{eq:0437L}). From left to right, in landscape orientation, the columns contain (1) $I/I_\text{max}$ (solid curve) and $L/I_\text{max}$ (dashed curve) profiles, plotted against pulse longitude (in units of degrees); (2) $I$-$Q$ phase portrait; (3) $I$-$U$ phase portrait; (4) $Q$-$U$ phase portrait, and (5) PA swing (dotted curve, in rad). }
\label{j0437all}
\end{figure*}

We now demonstrate how the Stokes phase portraits change as we add $I_2$ and $I_3$ to the filled core $I_1$. Figure \ref{j0437all} shows the pulse profile, PA swing, and phase portraits for $I_1$, $I_1 + I_2$, and $I_1 + I_2 + I_3$ respectively at an emission altitude of $r = 0.1 r_\text{LC}$ and $\beta = 90^\circ$.  

For just the filled core (Figure \ref{j0437all}, top row in landscape orientation), corresponding to peak C in the data, there is a thin, tilted balloon in $I$-$Q$, a balloon in $I$-$U$, and a tilted oval in $Q$-$U$. Aside from the obvious dissimilarity with the figure-eight seen in the data in $I$-$Q$, the oval in $Q$-$U$ does not resemble diagonal 1 in the data. The balloon in $I$-$U$ is an approximate match to the data, although it is not symmetric about $U = 0$.

 For the filled core and one hollow cone (Figure \ref{j0437all}, middle row), i.e. peaks B, C and D in the data, the conal components introduce a kink at $(I, Q) \approx (0.2, -0.03)$. In the $I$-$U$ plane, kinks are predicted to occur at $I \lesssim 0.2$. The data also contain kinks in this region. In the $Q$-$U$ plane, there is another kink near $(Q, U) \approx (0, -0.05)$, though the model is still a poor match to the data. 

The addition of the second hollow cone in Figure \ref{j0437all} (third row) completes the beam pattern. Still, the overall shapes of the phase portraits do not match the data.  The second cone appears as an extra kink in $I$-$Q$ and $I$-$U$ in the region $I \lesssim 0.15$. In $Q$-$U$, a secondary oval forms. At a stretch, it might be said that this secondary oval corresponds to one diagonal of an X-shape while the other large oval corresponds to another, but other interpretations are equally possible. 
Adjusting the emission altitude does not improve the fit.

\begin{figure}
\includegraphics[scale=0.4]{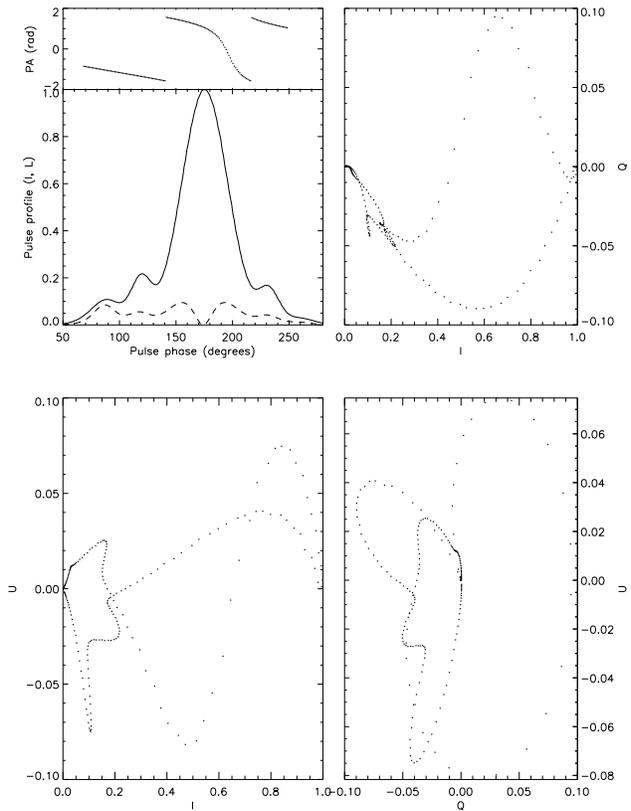}
\caption{Theoretical polarization model of PSR J0437$-$4715 for a pure dipole emitting at $r = 0.1 r_\text{LC}$ with $(\alpha, i) = (32^\circ, 26^\circ)$, beam pattern given by (\ref{eq:0437I1}), (\ref{eq:0437I2}) and (\ref{eq:0437I3}), and linear polarization given by (\ref{eq:0437L}). Clockwise from top left panel: (a) $I/I_\text{max}$ (lower subpanel, solid curve) and $L$ (lower subpanel, dashed curve) profiles, and PA swing (upper subpanel, dotted curve, in rad) all plotted against pulse longitude (in units of degrees) (b) $I$-$Q$ phase portrait; (c) $Q$-$U$ phase portrait; (d) $I$-$U$ phase portrait.}
\label{j0437dipoleall}
\end{figure}

For completeness, in Figure \ref{j0437dipoleall}, we present the pulse profile, PA swing, and phase portraits for a filled core with two hollow cones for a pure dipole field at the same orientation and altitude, and $\beta = 0$. The phase portraits are also a poor match to the data. The $I$-$Q$ plane features a tilted, asymmetric balloon, whereas the $I$-$U$ plane features an asymmetric figure-eight. In the $Q$-$U$ plane, there is a distorted oval surrounded by a tilted heart shape. Again, the basis-independent $Q$-$U$ shape does not resemble the data at all. The kinks seen in the phase portraits of the current-modified dipole also appear in the pure dipole. 

\begin{figure}
\includegraphics[scale=0.4]{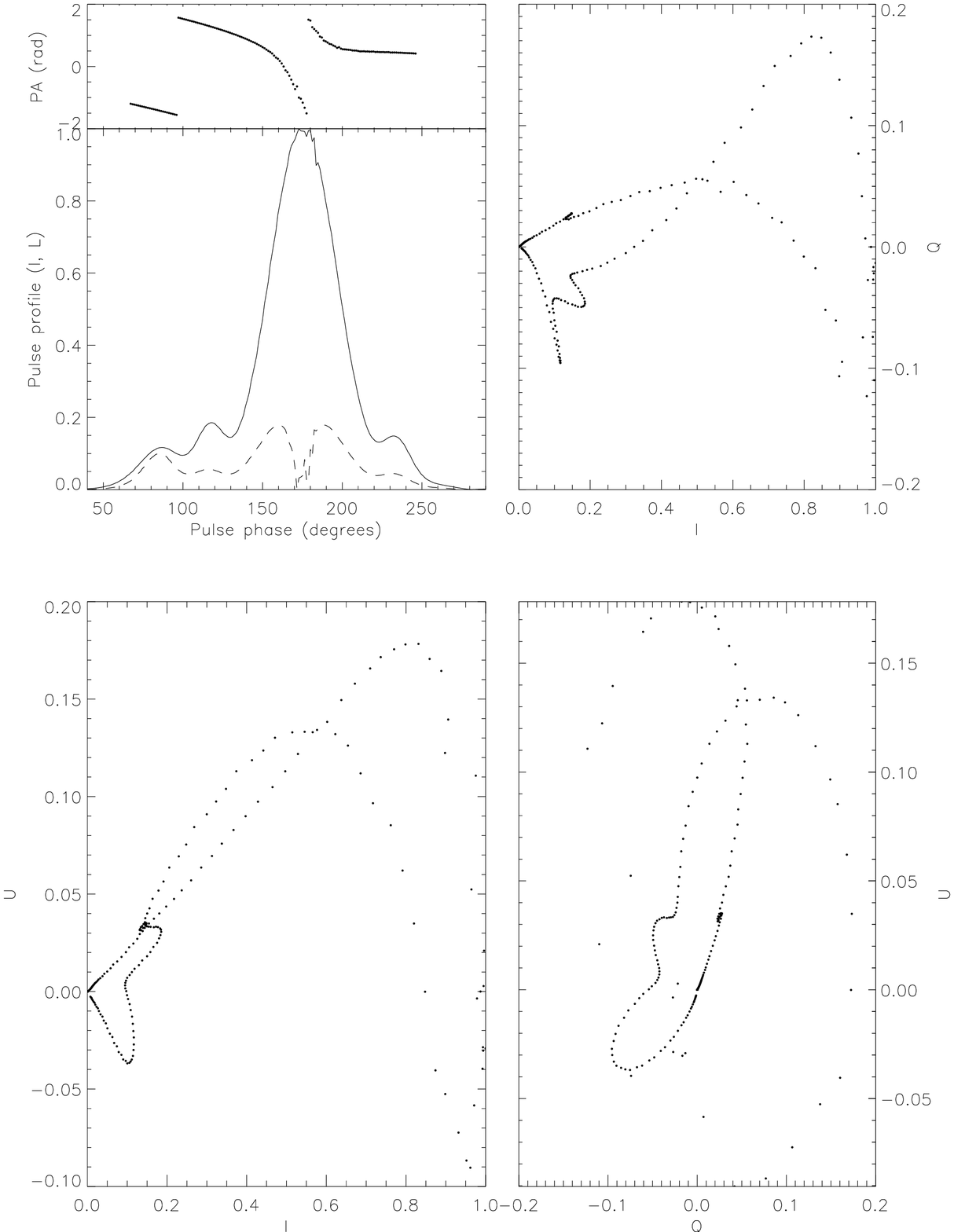}
\caption{Theoretical polarization model of PSR J0437$-$4715 for a simplified current-modified dipole given by (\ref{eq:simptoroid}) emitting at $r = 0.1 r_\text{LC}$ with $(\alpha, i) = (32^\circ, 26^\circ)$, beam pattern given by (\ref{eq:0437I1}), (\ref{eq:0437I2}) and (\ref{eq:0437I3}), and linear polarization given by (\ref{eq:0437L}). Clockwise from top left panel: (a) $I/I_\text{max}$ (lower subpanel, solid curve) and $L$ (lower subpanel, dashed curve) profiles, and PA swing (upper subpanel, dotted curve, in rad) all plotted against pulse longitude (in units of degrees) (b) $I$-$Q$ phase portrait; (c) $Q$-$U$ phase portrait; (d) $I$-$U$ phase portrait.}
\label{j0437simptoroidall}
\end{figure}

Finally, in Figure \ref{j0437simptoroidall} we present phase portraits for a simplified version of the current-modified field where the toroidal field is given by
\begin{equation}
\label{eq:simptoroid}
B_\phi = -B_p r/r_\text{LC}.
\end{equation}
In this stripped-down expression, $B_\phi$ depends on $\theta$ and $\alpha$ only through the poloidal field, and scales simply as $r/r_\text{LC}$. The phase portraits in FIgure \ref{j0437simptoroidall} are also presented for a filled core with two hollow cones at the same orientation and altitude, and $\beta = 0$. Again, they are a poor match to the data, although the $I$-$Q$ plane now features a tilted figure-eight similar to the data. The $I$-$U$ plane features an asymmetric figure-eight, and the $Q$-$U$ plane features two interlocking ovals. As in the previous cases, kinks corresponding to the various pulse peaks punctuate the phase portraits.

We note that, for the three magnetic geometries considered, the theoretical PA swings are smooth and relatively flat and do not contain any of the kinks seen in the data. Attempts to rotate the $Q$-$U$ portraits of the pure and simplified current-modified dipole to yield a better fit, i.e. trialling several values of $\beta$, are also unsuccessful. For example, for $\beta = 150^\circ$ with the simplified current-modified dipole, diagonal 1 in the $Q$-$U$ plane aligns with the large, interlocking ovals in the model while diagonal 2 aligns with the smaller, third oval. However the $I$-$Q$ plane now features a large balloon symmetric about $Q = 0$, and the $I$-$U$ plane features a broad, tilted balloon, neither of which matches the data.

We conclude that we are unable to fit the Stokes phase portraits for PSR J0437$-$4715 satisfactorily using a pure or current-modified dipole and a wide range of trial-and-error models for the beam and linear polarization patterns. There are several possible reasons for this. (i) The beam and polarization patterns might be very different, e.g. two highly asymmetric, nested cones. (ii) If the observed emission does indeed originate from multiple altitudes \citep{gangadhara08}, we would be unable to reproduce the Stokes phase portraits even if our guesses for $\alpha, i, I(t)$ and $L(t)$ are correct. (iii) The magnetic field in the emission region is neither a pure nor a current-modified dipole (very likely).

\section{Conclusion}
\label{sec:conclusion}
In this paper, we generalize the Stokes tomography technique introduced by CM10 by adding interpulse emission. In Section \ref{sec:population}, we present the Stokes phase portraits of 15 MSPs from the EPN online archive. By comparing the data to the generalized look-up tables for a current-modified dipole in the Appendix, we are able to infer approximately the geometric orientations for five of the MSPs. This is an improvement on the PA swing and rotating vector model, which yield orientations for only two of the objects --- orientations which, it transpires, are inconsistent with the observed Stokes phase portraits. In Section \ref{sec:j1939}, we model PSR J1939+2134 in detail, obtaining a match for the data at 0.61\,GHz with a current-modified dipole for $(\alpha, i) = (22 \pm 2^\circ, 80 \pm 1^\circ)$ and $r = 0.4r_\text{LC}$. However, we are unable to reproduce the data at 1.414\,GHz for the same orientation at altitudes in the range $0.22r_\text{LC} \leq r \leq 0.35r_\text{LC}$. In Section \ref{sec:j0437}, we repeat the process with PSR J0437$-$4715. At 1.44\,GHz, even the closest-matching orientation, $(\alpha, i) = (32^\circ, 26^\circ)$ with $r = 0.1 r_\text{LC}$, does not reproduce the data satisfactorily. 

The results from Sections \ref{sec:j1939} and \ref{sec:j0437} indicate that, while pure or current-modified dipoles are effective models for non-recycled pulsars (CM10), MSPs are likely to have more complicated magnetic geometries. This is not surprising, as the accretion process can significantly distort a pulsar's magnetic field \citep{lai99, pamel04, lamb09}. Alternative magnetic configurations include a quadrupole or localized surface anomaly \citep{lai99, long08}, a force-free field \citep{spitkovsky06, bai09}, a vacuum-like field \citep{melatos97}, or a field distorted by the formation of a polar magnetic mountain \citep{pamel04, vigmel08}. 

We emphasize the utility of the Stokes phase portraits as a supplementary diagnostic tool for MSPs. The PA swing on its own is especially ambiguous when dealing with non-dipolar fields. 
Future work will focus on the role played by circular polarization in Stokes tomography, the longitudinal structure of vacuum and force-free magnetospheres, and the polarization signatures of magnetic mountains. These topics will form the subject of companion papers.

\section*{Acknowledgements}
Part of this research has made use of the data base of published pulse profiles and Stokes parameters maintained by the European Pulsar Network, available at: http://www.mpifr-bonn.mpg.de/pulsar/data/. CC acknowledges the support of an Australian Postgraduate Award and the Albert Shimmins Memorial Fund.

\clearpage

\section*{Appendix: Atlas of look-up tables of Stokes phase portraits}
\begin{figure*}
\includegraphics[scale=0.8]{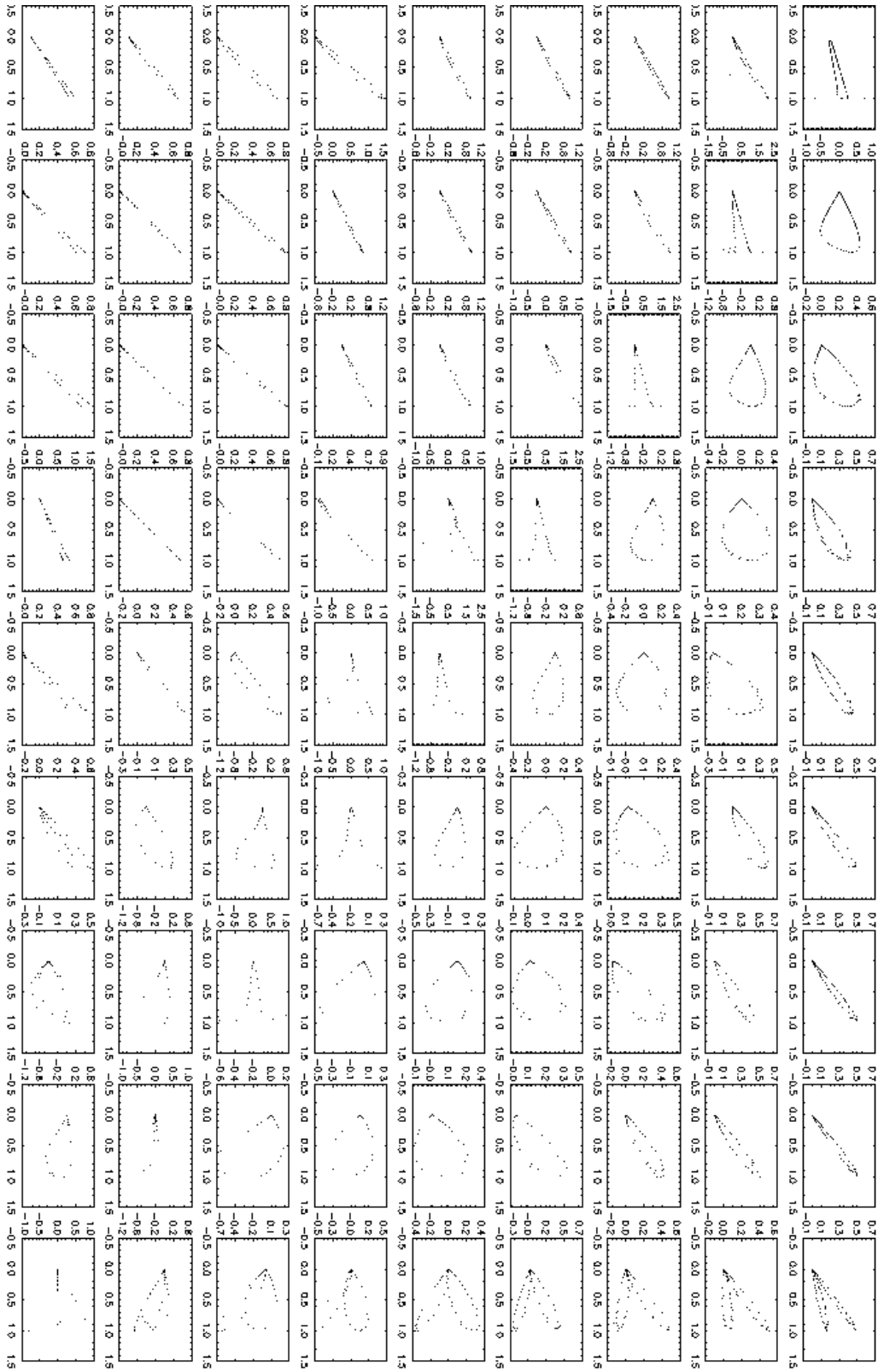}
\caption{Current-modified dipole. Look-up table of Stokes phase portraits in the $I$-$Q$ plane for filled core beams with degree of linear polarization $L = I \cos \theta$, where $\theta$ is the emission point colatitude, and $r = 0.1 r_\text{LC}$. The panels are organised in landscape mode, in order of increasing $10^\circ \leq i \leq 90^\circ$ (left--right) and $10^\circ \leq \alpha \leq 90^\circ$ (top--bottom) in intervals of $10^\circ$. $I$ is plotted on the horizontal axis and normalised by its peak value. $Q$ is plotted on the vertical axis.}
\label{m0w10lcostheta_tqvsi}
\end{figure*}

\begin{figure*}
\includegraphics[scale=0.8]{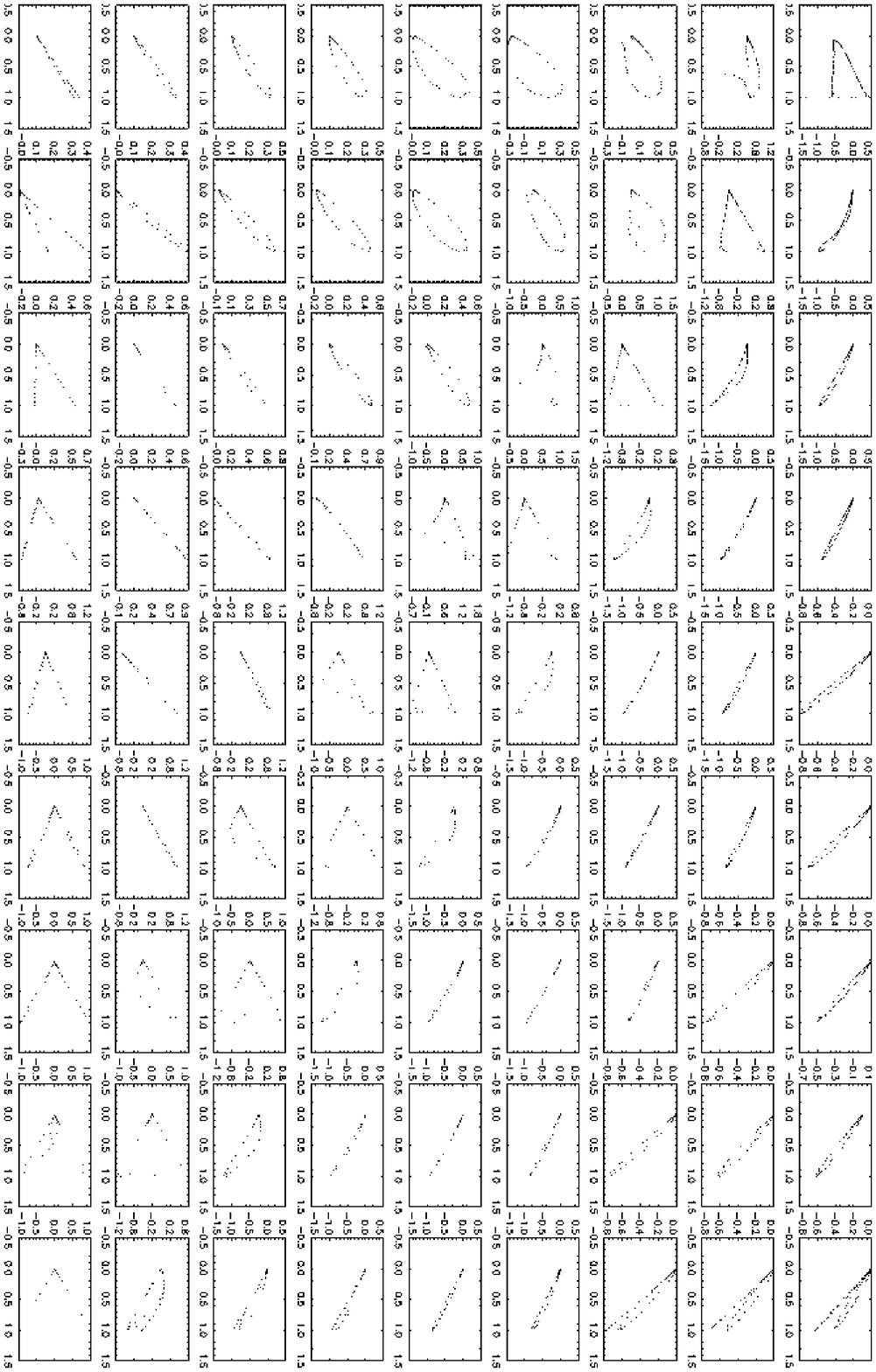}
\caption{Current-modified dipole. Layout as for Figure \ref{m0w10lcostheta_tqvsi}, but for $I$-$U$ ($I$ on the horizontal axis).}
\label{m0w10lcostheta_tuvsi}
\end{figure*}

\begin{figure*}
\includegraphics[scale=0.8]{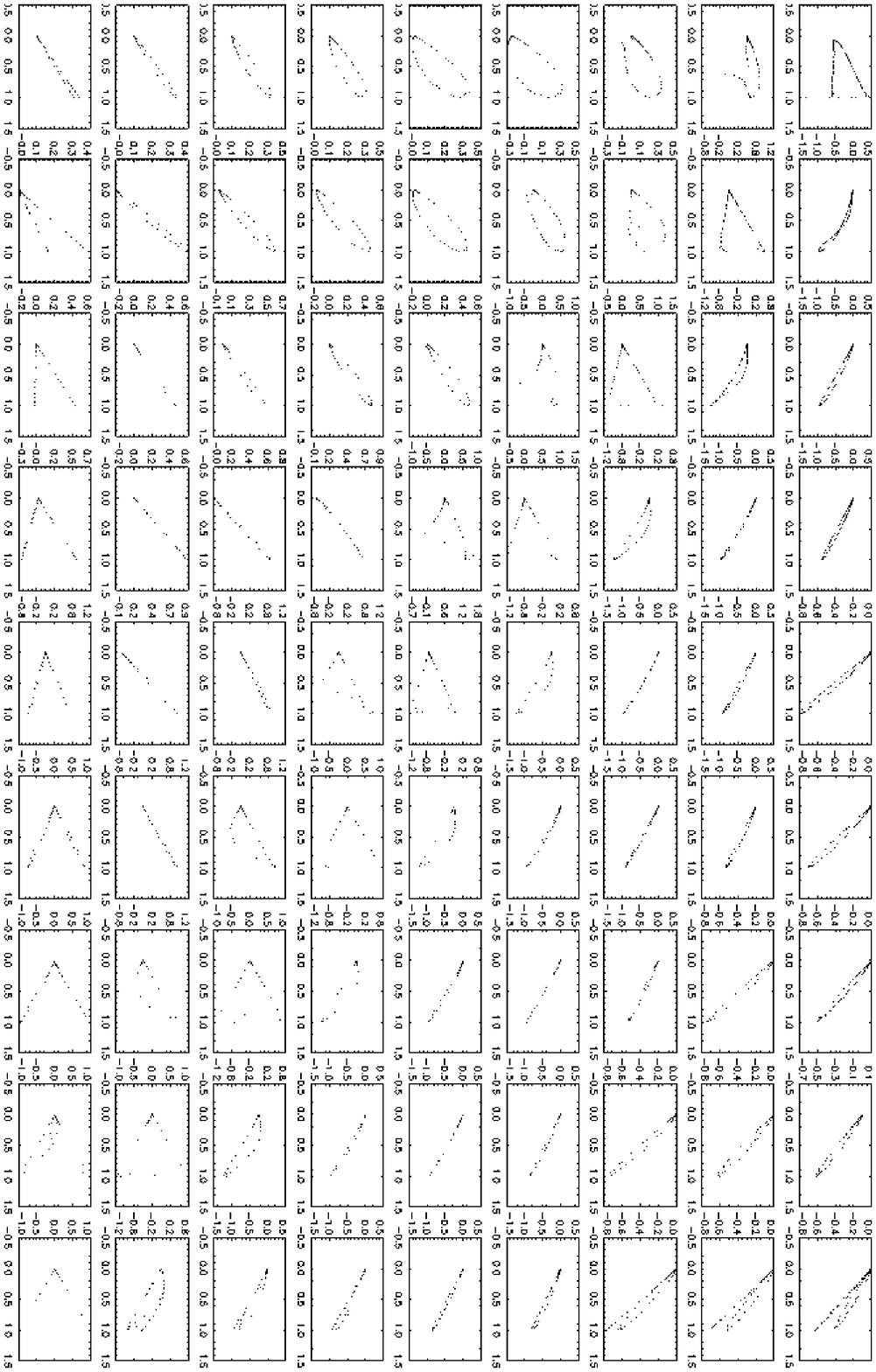}
\caption{Current-modified dipole. Layout as for Figure \ref{m0w10lcostheta_tqvsi}, but for $Q$-$U$ ($Q$ on the horizontal axis).}
\label{m0w10lcostheta_tuvsq}
\end{figure*}

\begin{figure*}
\includegraphics[scale=0.8]{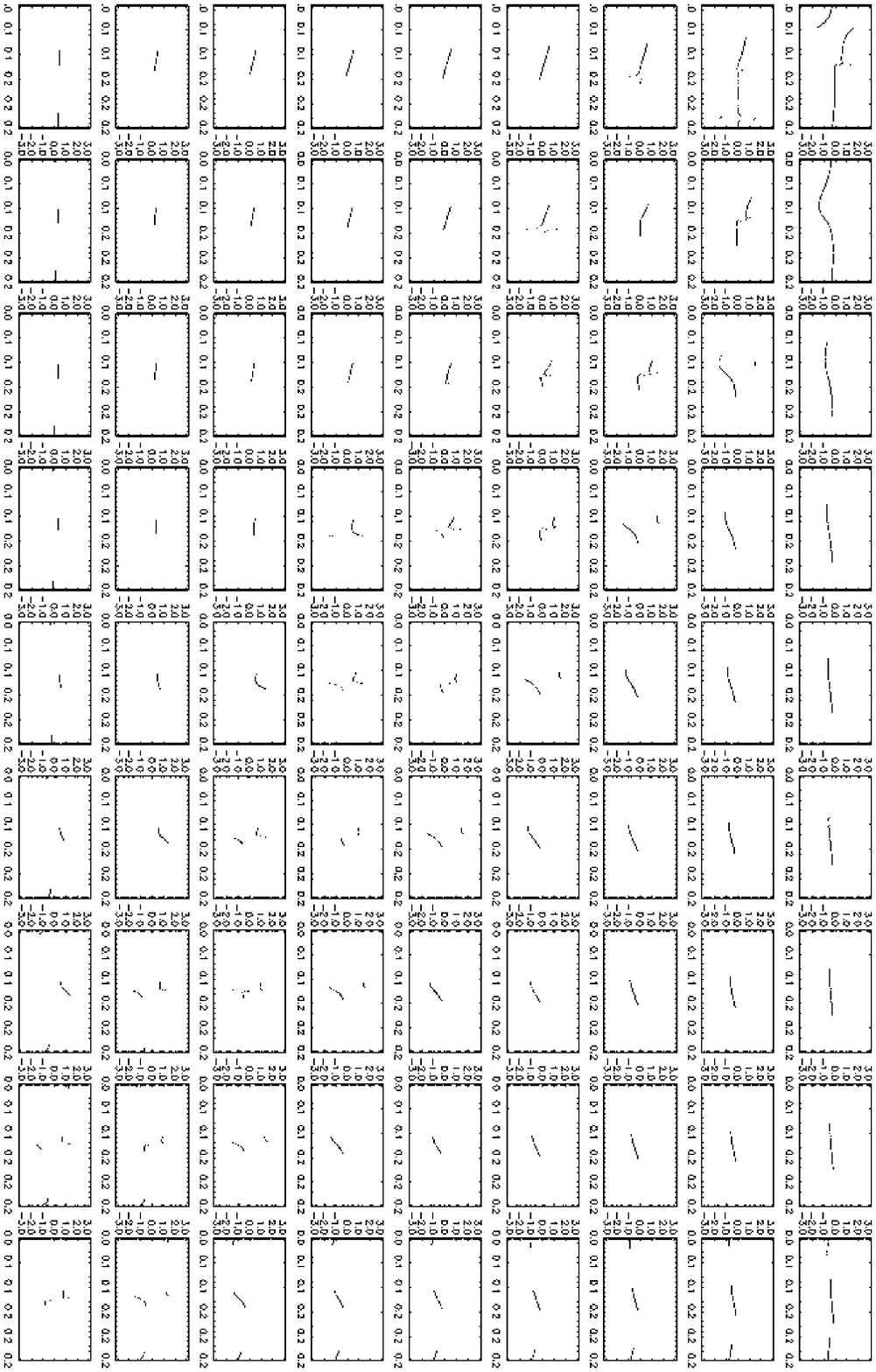}
\caption{Current-modified dipole. Layout as for Figure \ref{m0w10lcostheta_tqvsi}, but for position angle (on the vertical axis in landscape orientation, in units of radians) versus pulse longitude (on the horizontal axis, in units of $2 \pi$ radians).}
\label{m0w10lcostheta_tpa}
\end{figure*}

\begin{figure*}
\includegraphics[scale=0.8]{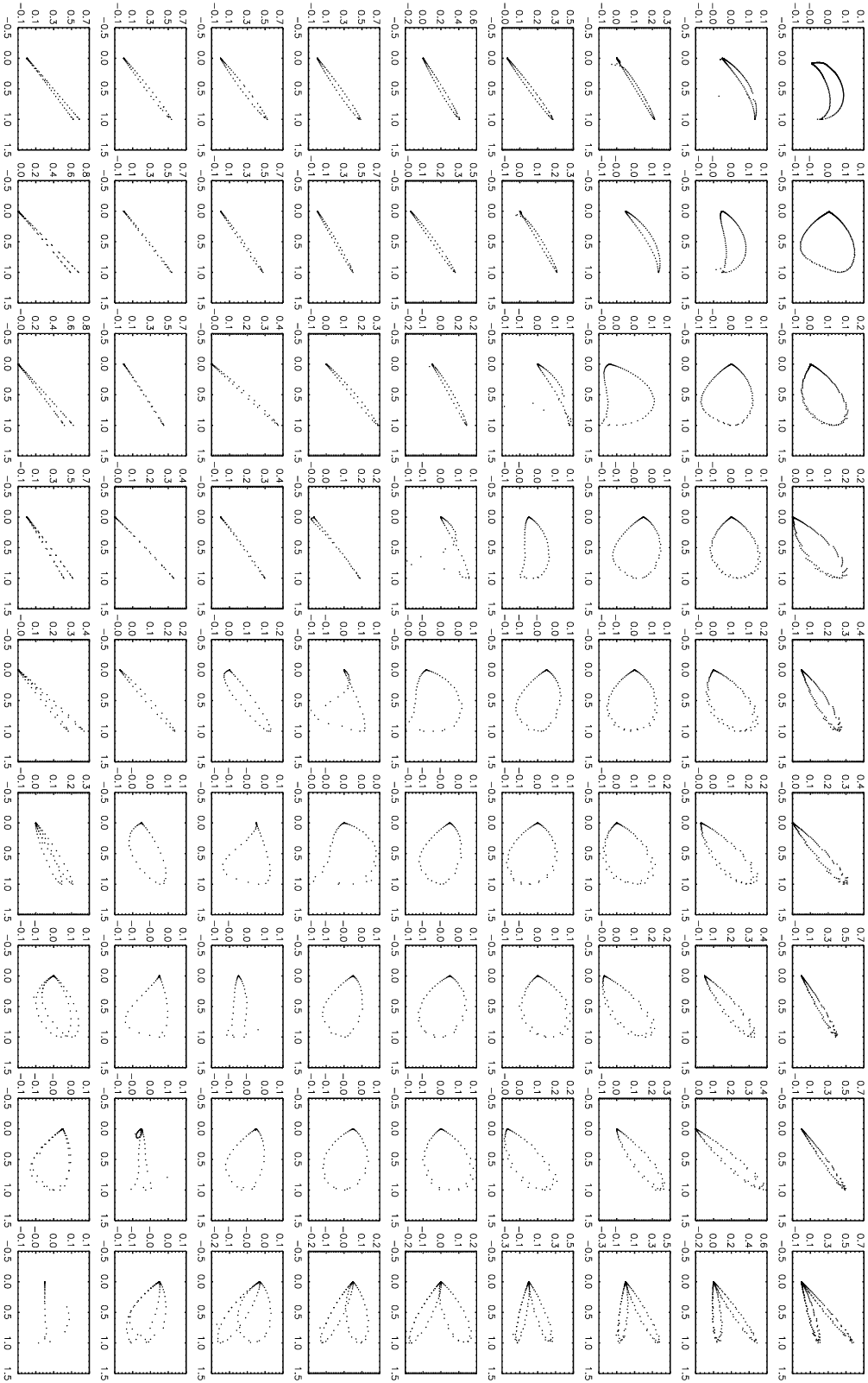}
\caption{Current-modified dipole. Look-up table of Stokes phase portraits in the $I$-$Q$ plane for filled core beams with degree of linear polarization $L = I \sin \theta$, where $\theta$ is the emission point colatitude, and $r = 0.1 r_\text{LC}$. The panels are organised in landscape mode, in order of increasing $10^\circ \leq i \leq 90^\circ$ (left--right) and $10^\circ \leq \alpha \leq 90^\circ$ (top--bottom) in intervals of $10^\circ$. $I$ is plotted on the horizontal axis and normalised by its peak value. $Q$ is plotted on the vertical axis.}
\label{m0w10lsintheta_tqvsi}
\end{figure*}

\begin{figure*}
\includegraphics[scale=0.8]{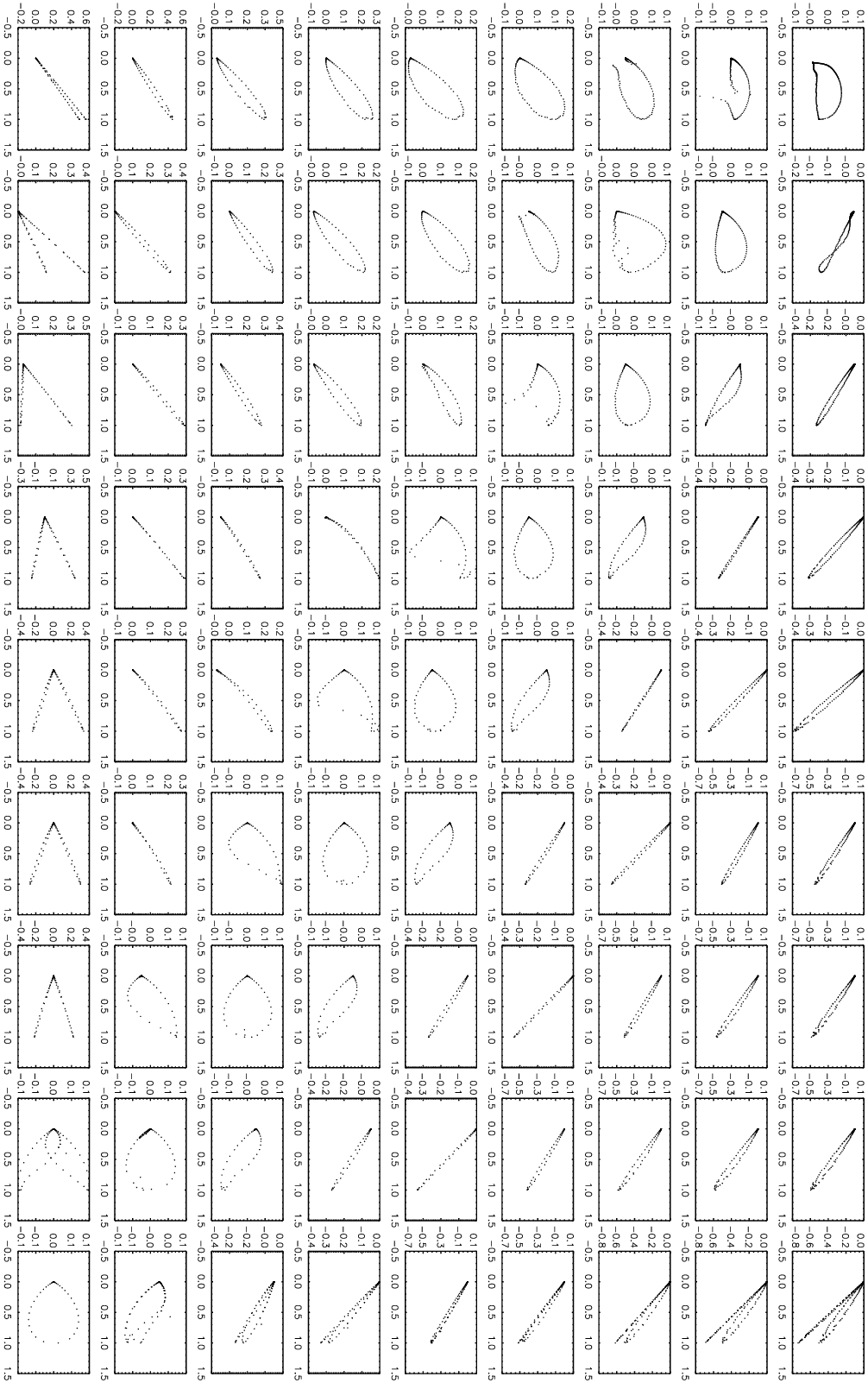}
\caption{Current-modified dipole. Layout as for Figure \ref{m0w10lsintheta_tqvsi}, but for $I$-$U$ ($I$ on the horizontal axis).}
\label{m0w10lsintheta_tuvsi}
\end{figure*}

\begin{figure*}
\includegraphics[scale=0.8]{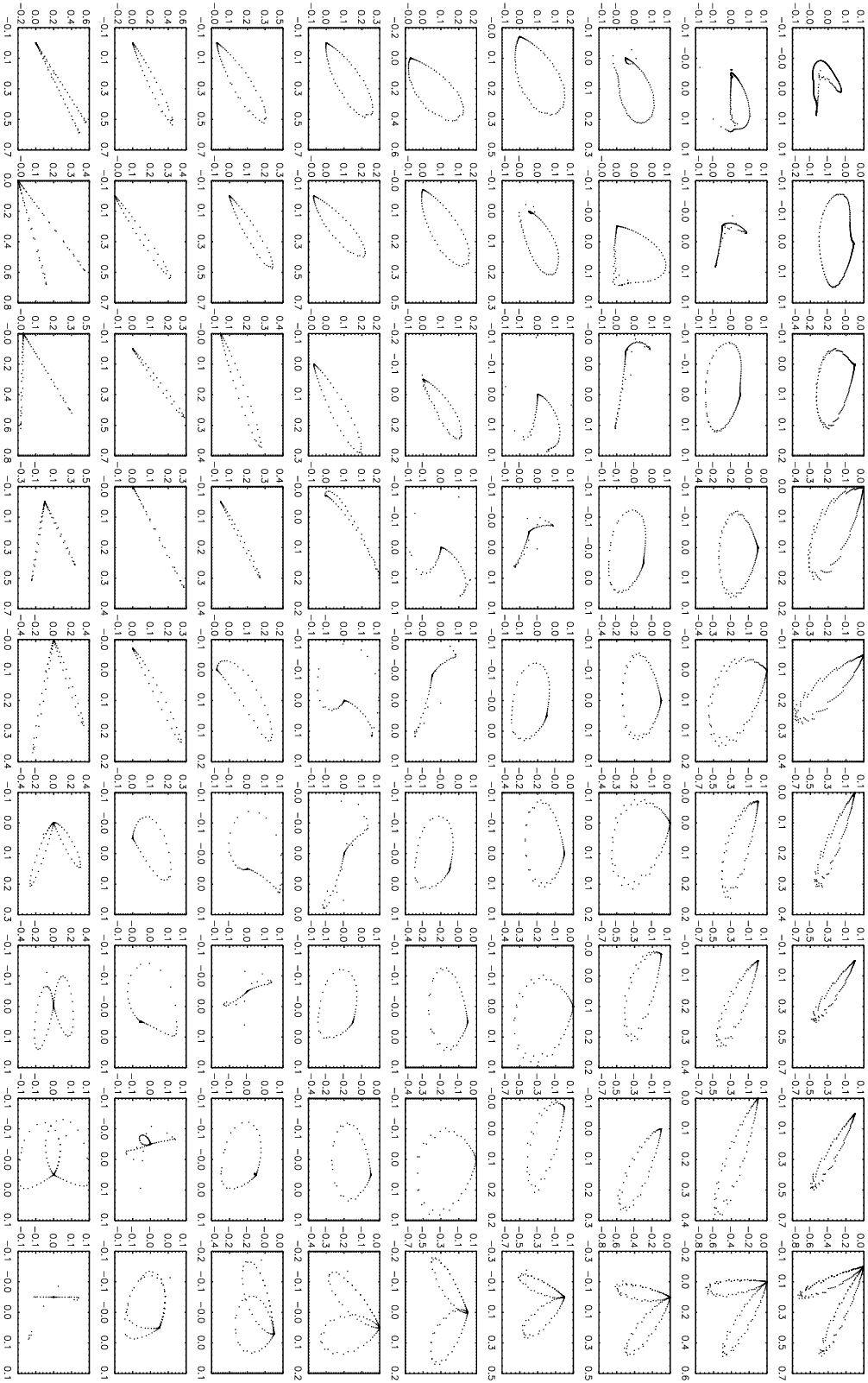}
\caption{Current-modified dipole. Layout as for Figure \ref{m0w10lsintheta_tqvsi}, but for $Q$-$U$ ($Q$ on the horizontal axis).}
\label{m0w10lsintheta_tuvsq}
\end{figure*}

\begin{figure*}
\includegraphics[scale=0.8]{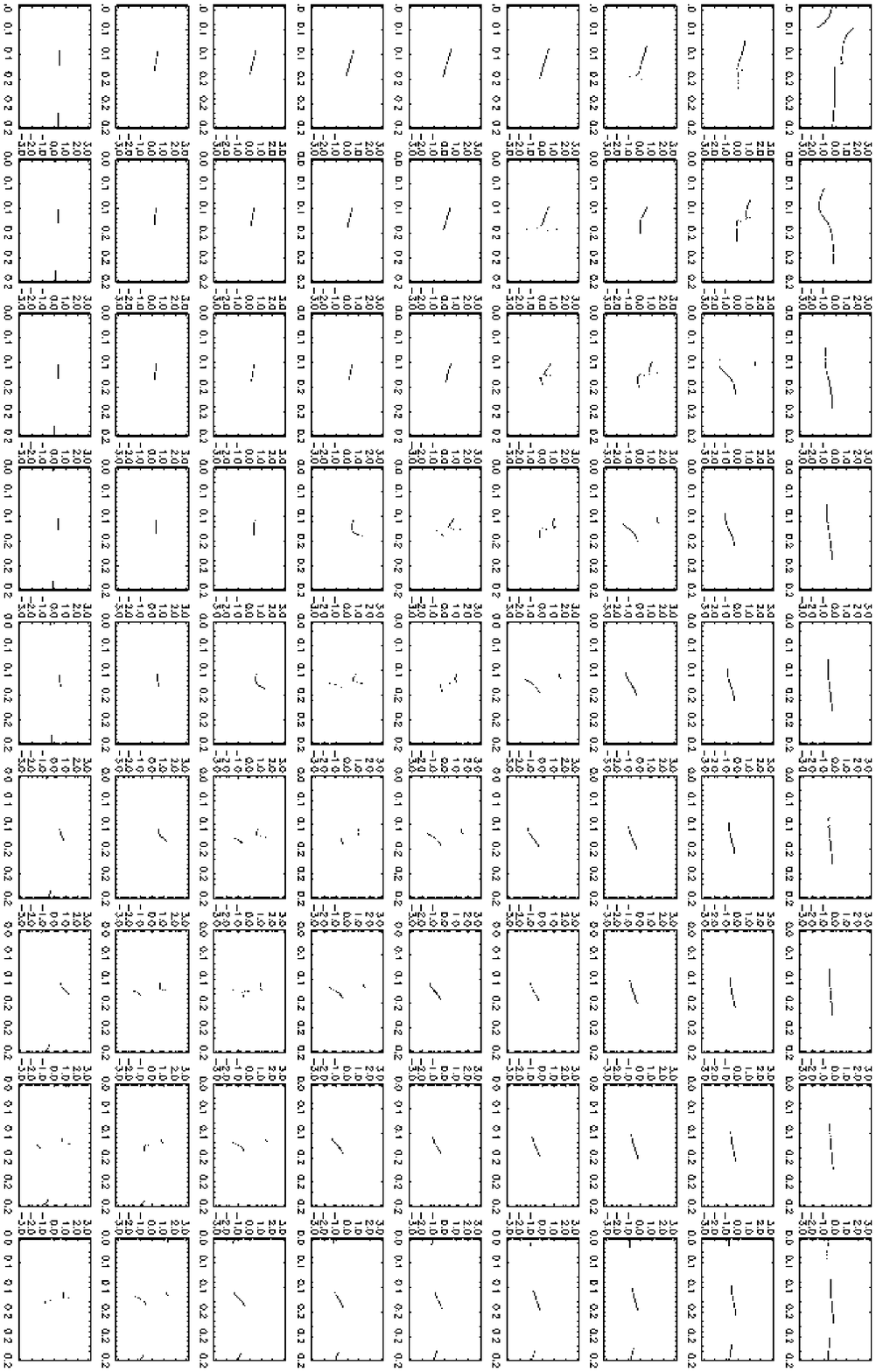}
\caption{Current-modified dipole. Layout as for Figure \ref{m0w10lsintheta_tqvsi}, but for position angle (on the vertical axis in landscape orientation, in units of radians) versus pulse longitude (on the horizontal axis, in units of $2 \pi$ radians).}
\label{m0w10lsintheta_tpa}
\end{figure*}

\begin{figure*}
\includegraphics[scale=0.8]{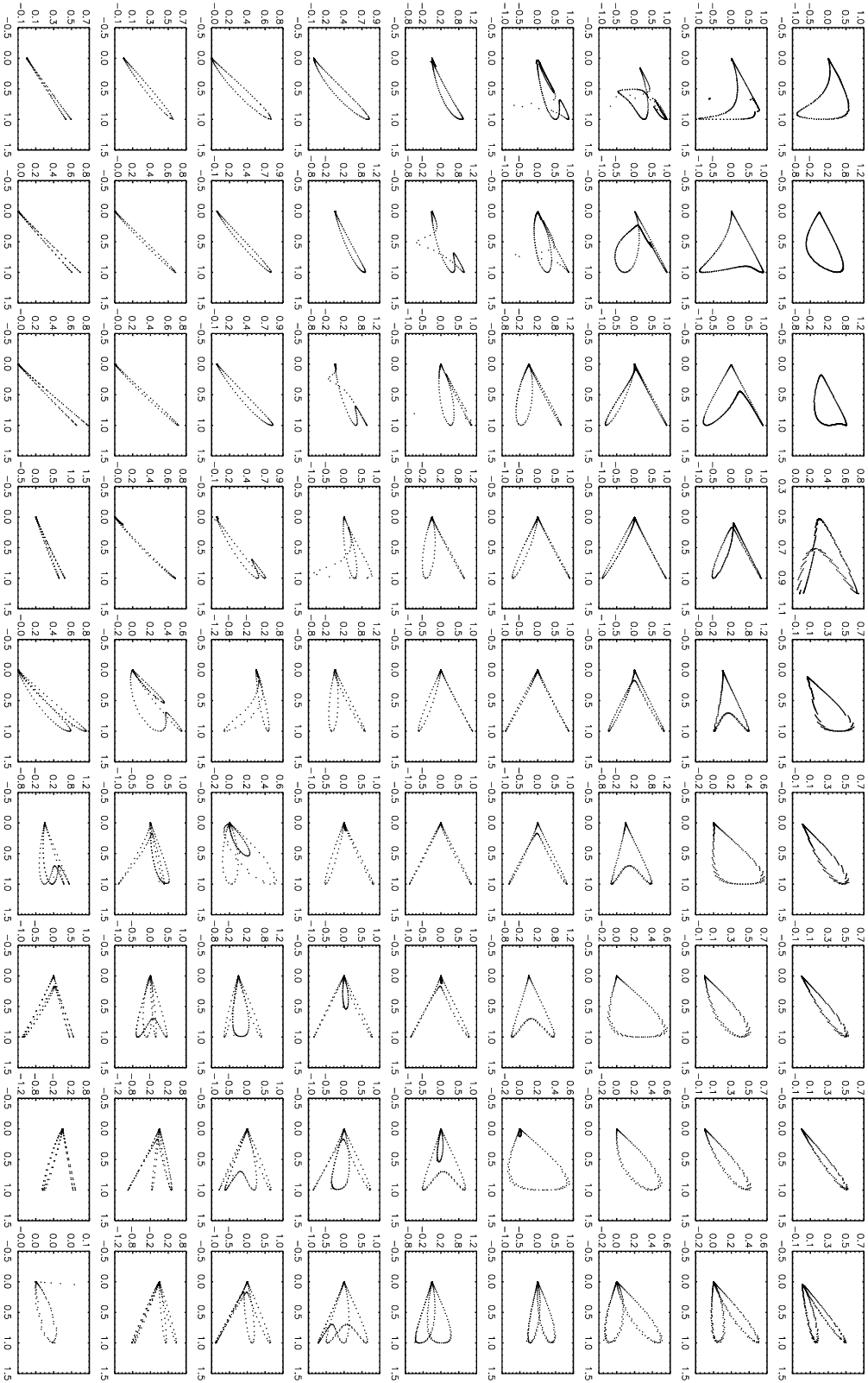}
\caption{Current-modified dipole. Look-up table of Stokes phase portraits in the $I$-$Q$ plane for hollow cones with opening angle $25^\circ$ and degree of linear polarization $L = I \cos \theta$, where $\theta$ is the emission point colatitude, and $r = 0.1 r_\text{LC}$. The panels are organised in landscape mode, in order of increasing $10^\circ \leq i \leq 90^\circ$ (left--right) and $10^\circ \leq \alpha \leq 90^\circ$ (top--bottom) in intervals of $10^\circ$. $I$ is plotted on the horizontal axis and normalised by its peak value. $Q$ is plotted on the vertical axis.}
\label{m25w10lcostheta_tqvsi}
\end{figure*}

\begin{figure*}
\includegraphics[scale=0.8]{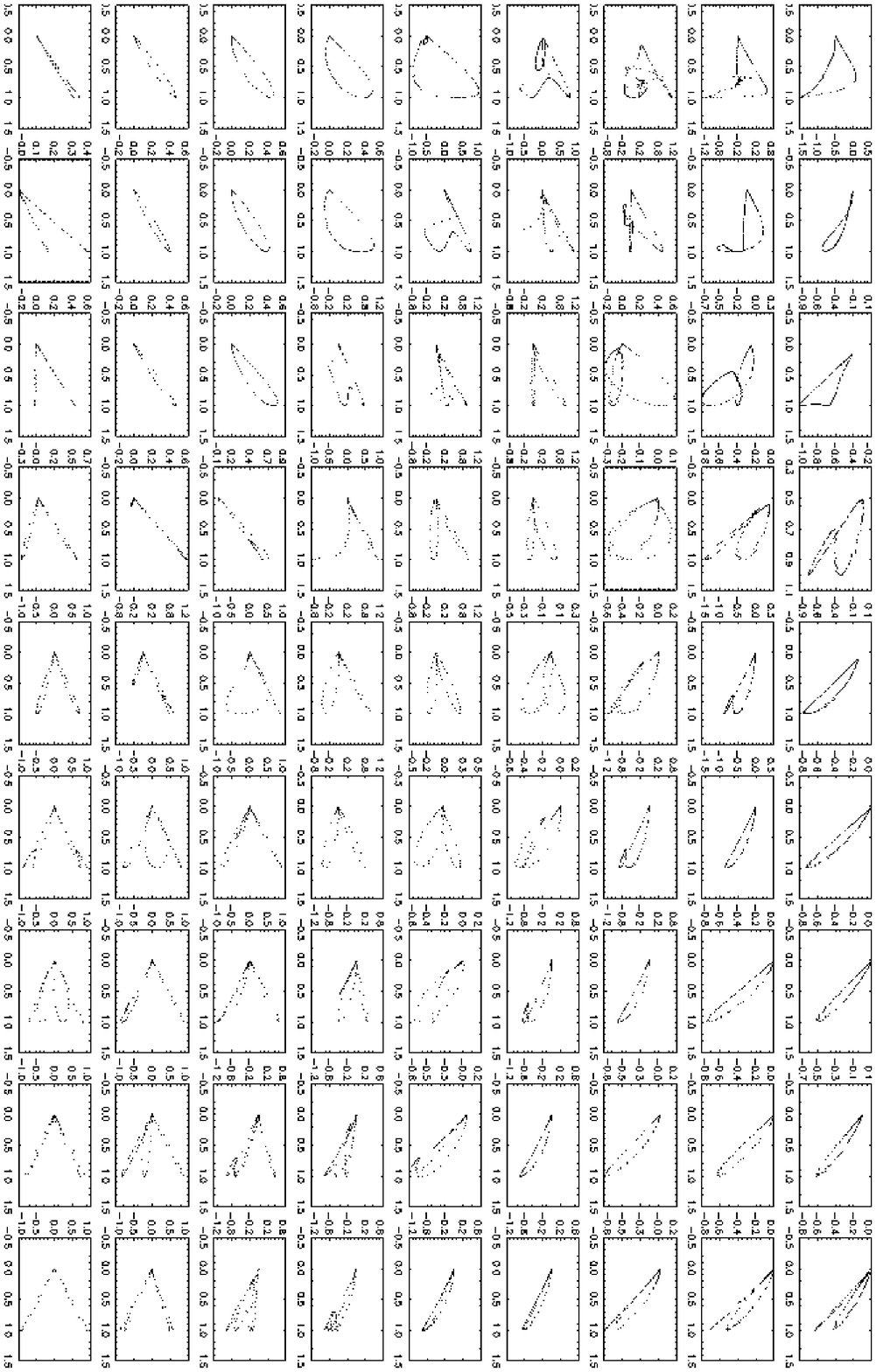}
\caption{Current-modified dipole. Layout as for Figure \ref{m25w10lcostheta_tqvsi}, but for $I$-$U$ ($I$ on the horizontal axis).}
\label{m25w10lcostheta_tuvsi}
\end{figure*}

\begin{figure*}
\includegraphics[scale=0.8]{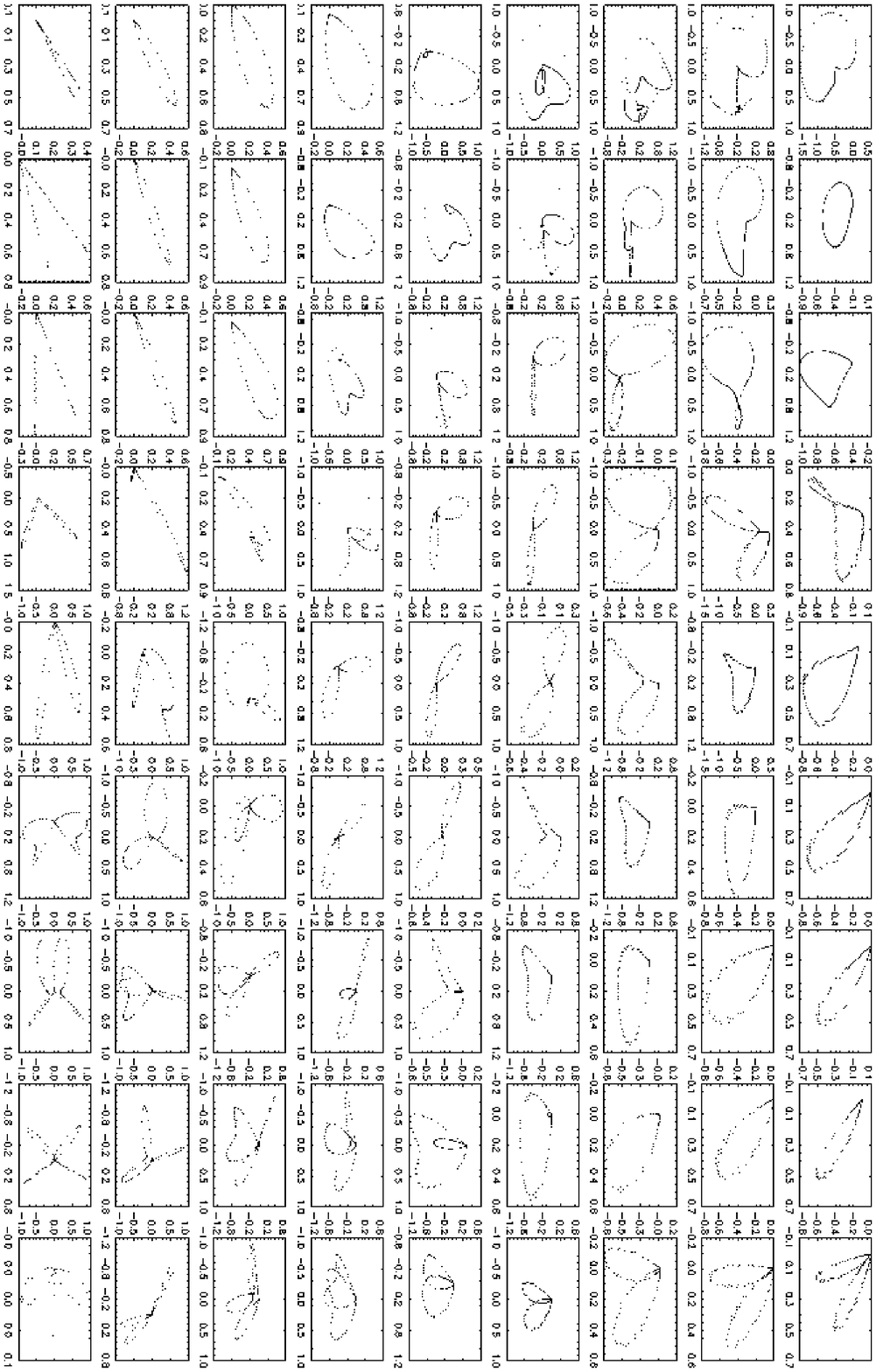}
\caption{Current-modified dipole. Layout as for Figure \ref{m25w10lcostheta_tqvsi}, but for $Q$-$U$ ($Q$ on the horizontal axis).}
\label{m25w10lcostheta_tuvsq}
\end{figure*}

\begin{figure*}
\includegraphics[scale=0.8]{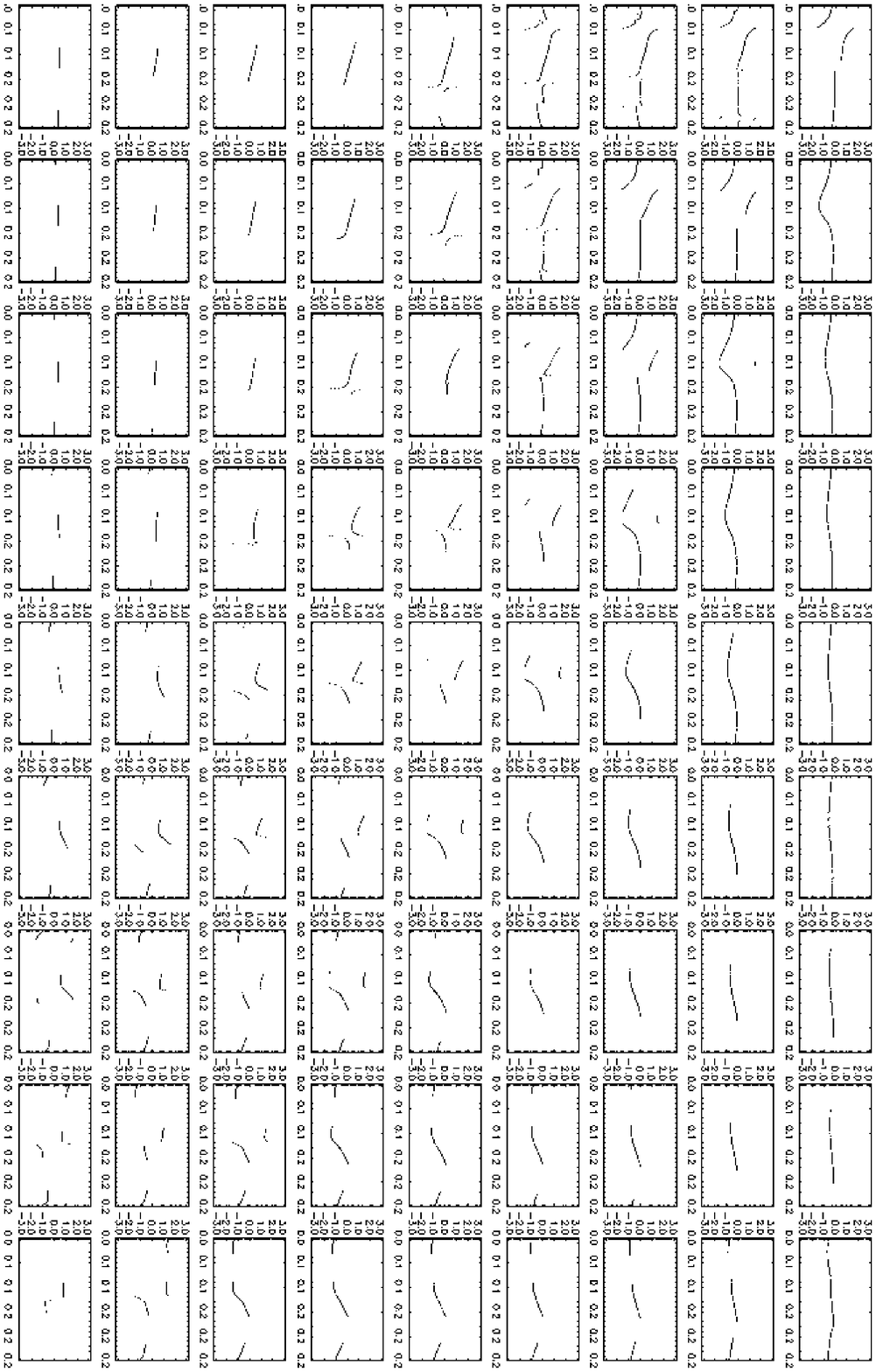}
\caption{Current-modified dipole. Layout as for Figure \ref{m25w10lcostheta_tqvsi}, but for position angle (on the vertical axis in landscape orientation, in units of radians) versus pulse longitude (on the horizontal axis, in units of $2 \pi$ radians).}
\label{m25w10lcostheta_tpa}
\end{figure*}

\begin{figure*}
\includegraphics[scale=0.8]{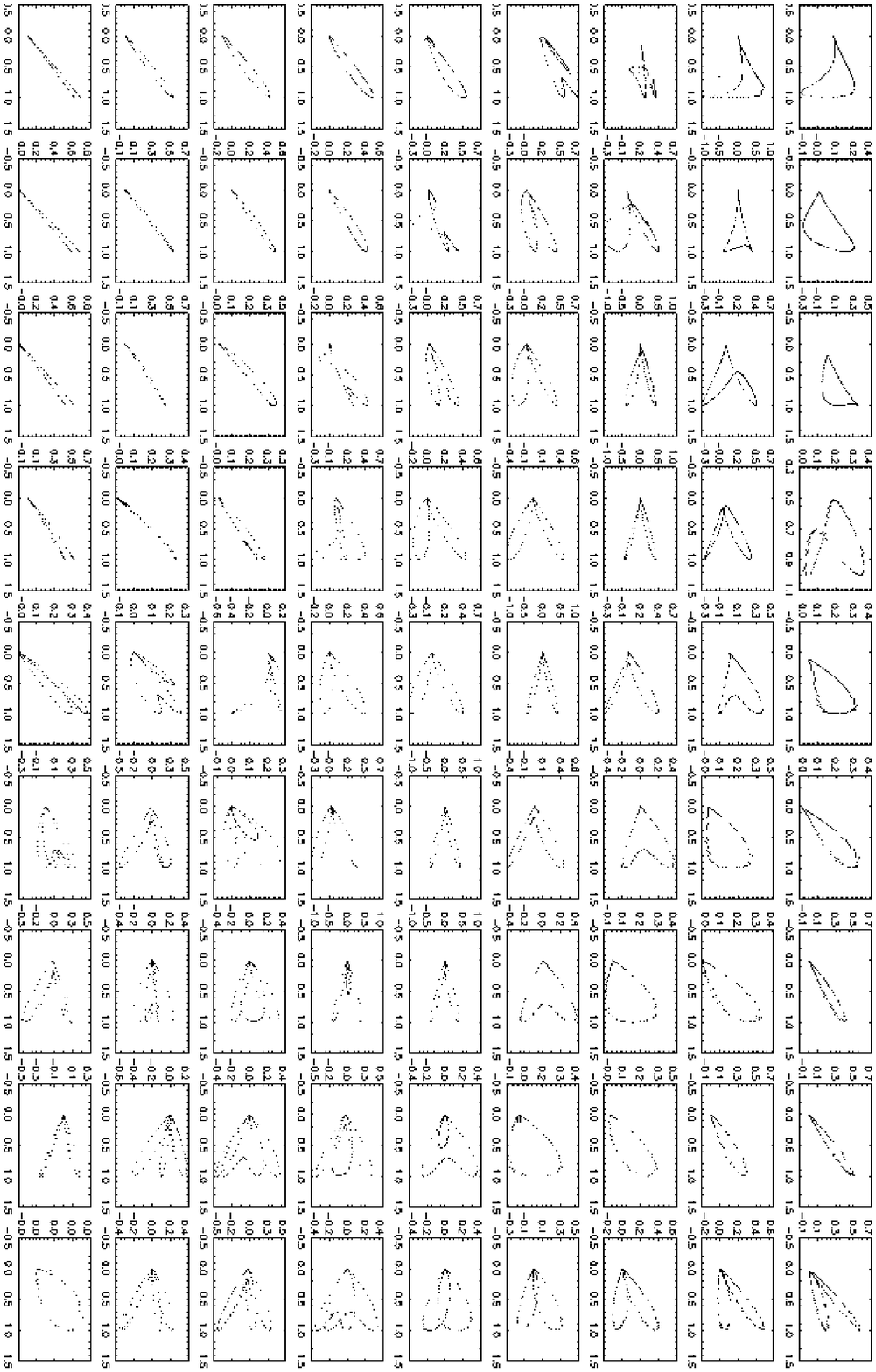}
\caption{Current-modified dipole. Look-up table of Stokes phase portraits in the $I$-$Q$ plane for hollow cones with opening angle $25^\circ$ and degree of linear polarization $L = I \sin \theta$, where $\theta$ is the emission point colatitude, and $r = 0.1 r_\text{LC}$. The panels are organised in landscape mode, in order of increasing $10^\circ \leq i \leq 90^\circ$ (left--right) and $10^\circ \leq \alpha \leq 90^\circ$ (top--bottom) in intervals of $10^\circ$. $I$ is plotted on the horizontal axis and normalised by its peak value. $Q$ is plotted on the vertical axis.}
\label{m25w10lsintheta_tqvsi}
\end{figure*}

\begin{figure*}
\includegraphics[scale=0.8]{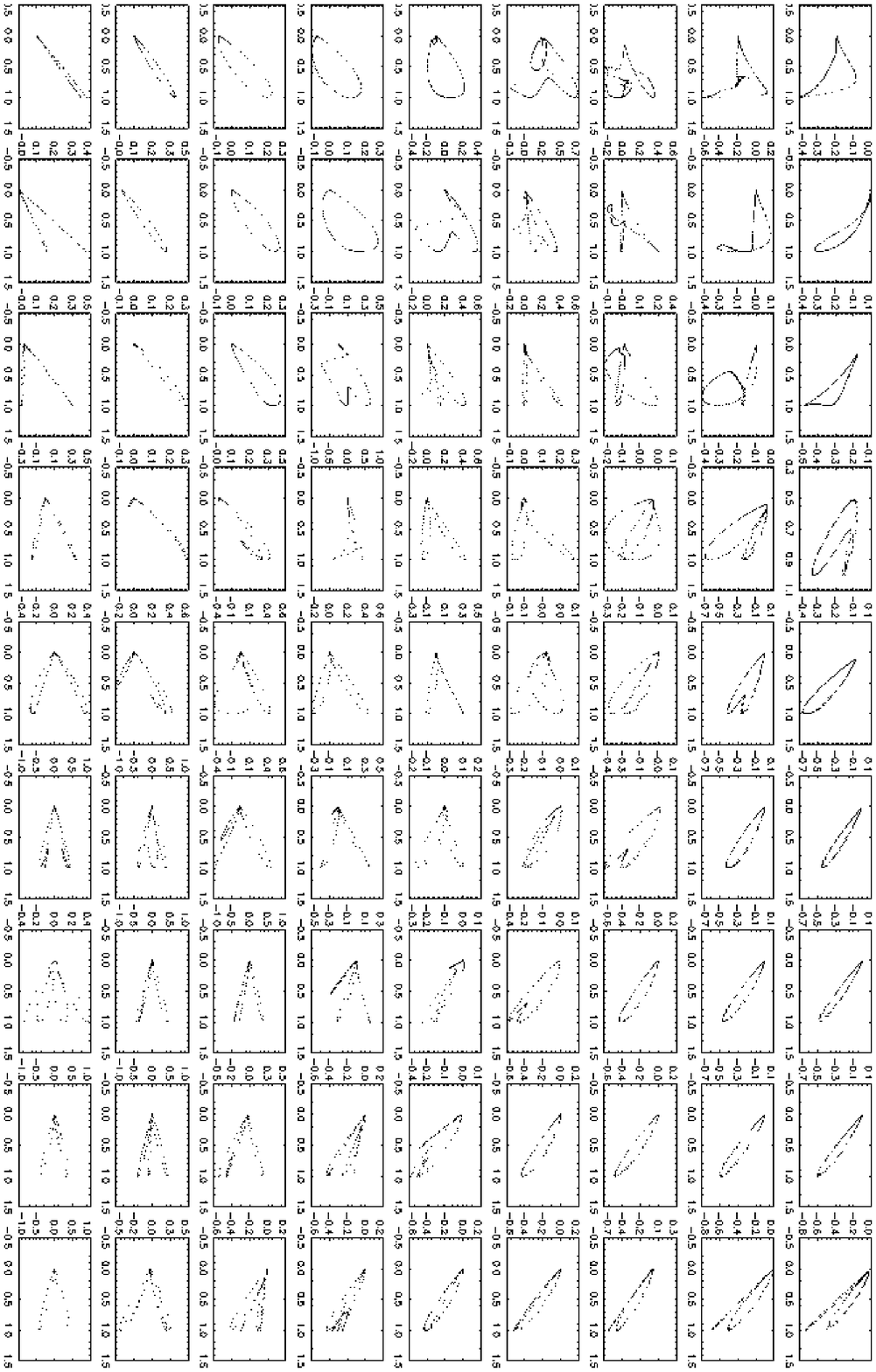}
\caption{Current-modified dipole. Layout as for Figure \ref{m25w10lsintheta_tqvsi}, but for $I$-$U$ ($I$ on the horizontal axis).}
\label{m25w10lsintheta_tuvsi}
\end{figure*}

\begin{figure*}
\includegraphics[scale=0.8]{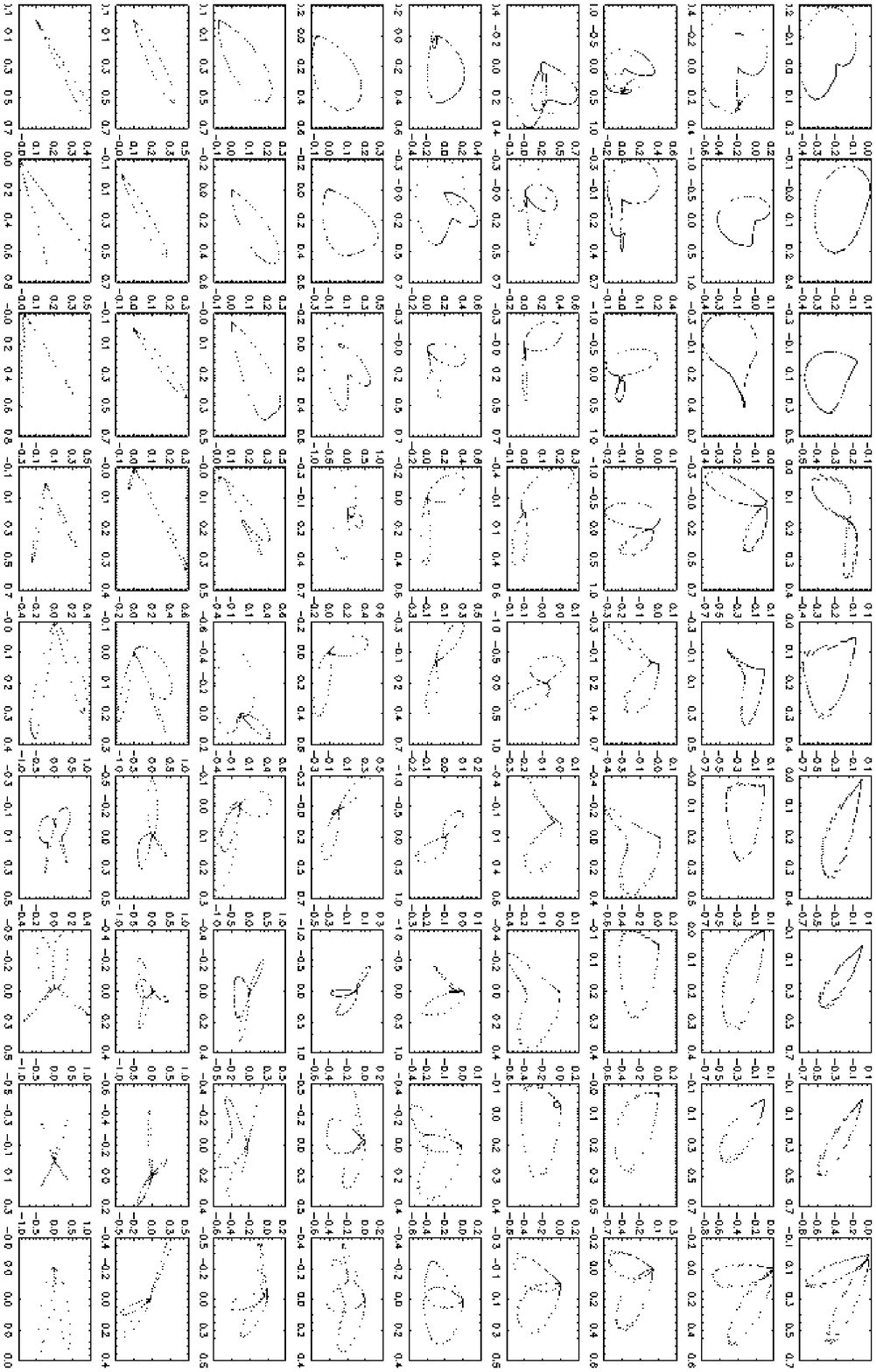}
\caption{Current-modified dipole. Layout as for Figure \ref{m25w10lsintheta_tqvsi}, but for $Q$-$U$ ($Q$ on the horizontal axis).}
\label{m25w10lsintheta_tuvsq}
\end{figure*}

\begin{figure*}
\includegraphics[scale=0.8]{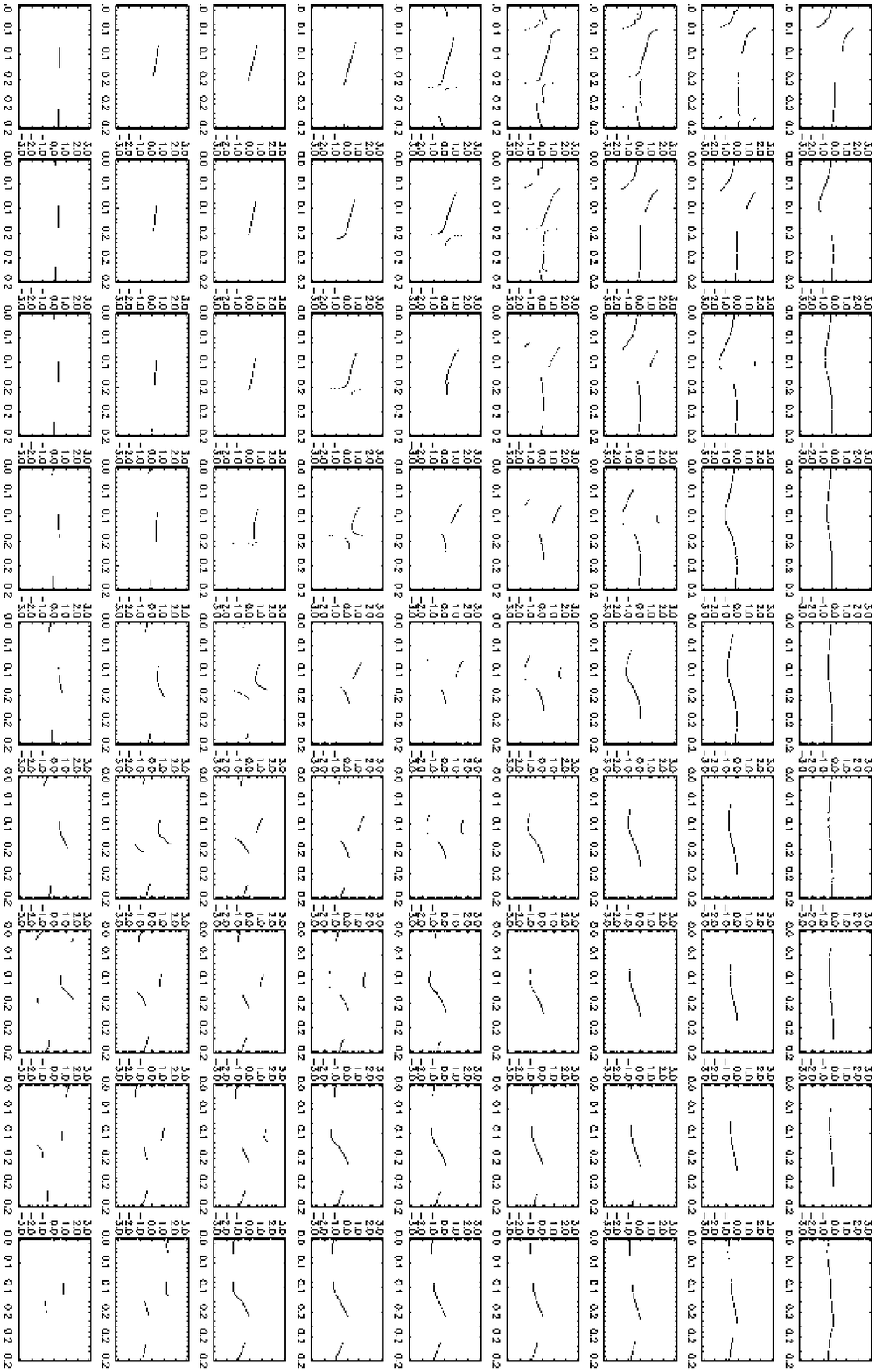}
\caption{Current-modified dipole. Layout as for Figure \ref{m25w10lsintheta_tqvsi}, but for position angle (on the vertical axis in landscape orientation, in units of radians) versus pulse longitude (on the horizontal axis, in units of $2 \pi$ radians).}
\label{m25w10lsintheta_tpa}
\end{figure*}

We present look-up tables for the beam patterns and linear polarization models used in CM10, updated to include interpulse and relativistic aberration effects. All figures are for $r = 0.1 r_\text{LC}$. Stokes phase portraits and PA swings are shown for a current-modified dipole with
\begin{enumerate}
\item a filled core beam with $L = I \cos \theta$ (Figures \ref{m0w10lcostheta_tqvsi}--\ref{m0w10lcostheta_tpa}),
\item a filled core beam with $L = I \sin \theta$ (Figures \ref{m0w10lsintheta_tqvsi}--\ref{m0w10lsintheta_tpa}),
\item a hollow cone with  $L = I \cos \theta$ (Figures \ref{m25w10lcostheta_tqvsi}--\ref{m25w10lcostheta_tpa}), and
\item a hollow cone with  $L = I \sin \theta$ (Figures \ref{m25w10lsintheta_tqvsi}--\ref{m25w10lsintheta_tpa}).
\end{enumerate}

\clearpage

\bibliographystyle{mn2e}
%\bibliography{AA_mnemonic,mspprofiles}

\label{lastpage}
\end{document}